\documentclass[preprint]{elsarticle}

\usepackage{graphicx,rotating}
\usepackage{epsfig}
\usepackage{amssymb}

\journal{Nuclear Physics A}

\newcommand{\ba}{\begin{eqnarray}}
\newcommand{\ea}{\end{eqnarray}}

\input{tcilatex}

\begin{document}

\begin{frontmatter}

\title{Cluster structure of $^{21}$Ne and $^{21}$Na}
\author{R. Bijker}
\address{Instituto de Ciencias Nucleares, 
Universidad Nacional Aut\'onoma de M\'exico, \\
Apartado Postal 70-543, 04510 Cd. de M\'exico, M\'exico}
\ead{bijker@nucleares.unam.mx}

\author{F. Iachello}
\address{Center for Theoretical Physics, Sloane Laboratory, Yale University, \\ 
New Haven, CT 06520-8120, U.S.A.}
\ead{francesco.iachello@yale.edu}

\begin{abstract}
We study the cluster structure of $^{21}$Ne and $^{21}$Na within the framework
of the cluster shell model (CSM) and show that they have a complex cluster
structure with the coexistence of a $^{20}$Ne+n, $^{20}$Ne+p structure and a 
$^{19}$Ne+2n, $^{19}$F+2p structure. Seven rotational bands are identified
in $^{21}$Ne and four in $^{21}$Na and assigned to single-particle cluster
states, single-hole cluster states and vibrational states. The single-particle 
states are associated with the $^{20}$Ne+n and $^{20}$Ne+p cluster structure, 
while the single-hole states are associated with the $^{19}$Ne+2n
and $^{19}$F+2p structure.
\end{abstract}


\end{frontmatter}

\section{Introduction}

Cluster structures in light nuclei were suggested in 1965 by Brink \cite%
{brink} and later extensively investigated within the framework of the
Brink-Bloch model \cite{brink1,weiguny1,weiguny2} for all
nuclei with $N=Z$ composed of $k$ $\alpha $-particles to be denoted
henceforth by $k\alpha $ nuclei. Several geometric configurations were
analyzed and it was concluded that the geometric structure of the ground
state of $^{8}$Be ($k=2$) is a dumbbell (${\cal Z}_{2}$ symmetry), of $^{12}$C ($%
k=3$) is a triangle (${\cal D}_{3h}$ symmetry) and of $^{16}$O ($k=4$) is a
tetrahedron (${\cal T}_{d}$ symmetry). These three structures have been recently
re-analyzed \cite{bijker1,bijker2,bijker3,bijker4,dellarocca2}, especially 
in $^{12}$C, in view of recent experimental
data on the rotational structure of the ground state band and of the
so-called Hoyle band \cite{freer1,freer2,freer3,gai}. In the Brink model 
\cite{brink1}, also $^{20}$Ne ($k=5$), $^{24}$Mg ($k=6$%
), and $^{28}$Si ($k=7$) were analyzed with suggested configurations for $%
^{20}$Ne of a bi-pyramid (${\cal D}_{3h}$ symmetry), for $^{24}$Mg a rhombic
bi-pyramid (${\cal D}_{2h}$ symmetry) and for $^{28}$Si a double-winged structure
with either ${\cal D}_{2h}$ or ${\cal D}_{2d}$ symmetry. For these nuclei several other structures have been suggested, in particular for $^{20}$Ne a body-centered
distorted tetrahedron \cite{hauge} and a $^{16}$O+$\alpha $ structure \cite%
{vonoertzen}. Very recently, we have provided evidence \cite{bijker5}, on
the basis of an extensive analysis of data accumulated in the last 50 years,
that the cluster structure of $^{20}$Ne appears to be a bi-pyramid thus
confirming the suggestion of Brink \cite{brink1}. 

An important question is the extent to which cluster structures survive 
the addition or subtraction of particles. We
denote these structures by $k\alpha +x$ nuclei, where $x=1,2,...$, is the
number of additional particles. These structures were originally suggested
by von Oertzen \cite{vonoertzen1,vonoertzen2,vonoertzen3,vonoertzen4}. 
The simplest case is $x=1$. In a previous publication,
one of us (F.I.) with Della Rocca analyzed the structure of $^{9}$Be and $%
^{9}$B as $^{8}$Be+n and $^{8}$Be+p \cite{dellarocca2} by using the
so-called cluster shell model (CSM) \cite{dellarocca1,santana}. 
In another publication we analyzed the structure of $^{13}$C as $^{12}$C+n \cite{bijker6}. In view of recent experimental
data in $^{21}$Ne \cite{wheldon} combined with old measurements \cite%
{firestone}, we analyze in this paper the cluster structure of $^{21}$Ne and
show that this nucleus has a complex structure with the coexistence of
particle and hole states with cluster structure $^{20}$Ne+n and $^{19}$%
Ne+2n. We also analyze the structure of the mirror nucleus $^{21}$Na and
show that it can be described by the coexistent cluster structures $^{20}$%
Ne+p and $^{19}$F+2p. The approach of this paper generalizes the previous
approach to $k\alpha$+x nuclei \cite{dellarocca2,dellarocca1} to
more complex situations.

In a major development of the last few years, large scale shell model
calculations have become feasible. In particular the development of the SDPF
interaction \cite{caurier} has allowed large scale calculations of s-d shell
nuclei. The analysis of experimental data presented here gives the
opportunity to test whether cluster features can be obtained from large
shell model calculations. A comparison between shell model and cluster
interpretation will be presented elsewhere \cite{mengoni}.

The structure of this paper is as follows. In Sect.~\ref{sec2}, we briefly review the
cluster structure of $^{20}$Ne. In Sect.~\ref{sec3}, we assign states in $^{21}$Ne to
rotational bands. These assignments are for most cases identical to those
given in Ref.~\cite{wheldon}. However, differences occur for two bands which have
a major effect on the interpretation of the cluster structure of $^{21}$Ne.
In Sect.~\ref{sec4}, we provide an interpretation of the observed rotational bands in
terms of the cluster shell model (CSM) \cite{dellarocca2}, summarize our
assignments, interpret those in terms of clustering and compare with other
assignments and models. In Sect.~\ref{sec5}, we analyze the available experimental
information in $^{21}$Ne on electromagnetic transition rates and moments and
compare with cluster calculations. In Sect.~\ref{sec6}, we discuss the structure of $^{21}$Na and compare it with that of $^{21}$Ne. Finally, Sect.~\ref{sec7} contains a summary and conclusions.

\section{Cluster structure of $^{20}$Ne}
\label{sec2}

\begin{figure}
\centering
\vspace{15pt}
\setlength{\unitlength}{0.75pt}
\begin{picture}(200,350)(0,0)
\thicklines
\put( 60,120) {\circle*{10}} 
\put( 30,150) {\circle*{10}}
\put(150,150) {\circle*{10}}
\put( 90,270) {\circle*{10}}
\put( 90, 30) {\circle*{10}}
\multiput( 30,150)(4,0){30}{\circle*{1}}
\put( 30,150) {\line( 1, 2){60}}
\put( 30,150) {\line( 1,-2){60}}
\put( 30,150) {\line( 1,-1){30}}
\put(150,150) {\line(-1, 2){60}}
\put(150,150) {\line(-1,-2){60}}
\put(150,150) {\line(-3,-1){90}}
\put( 60,120) {\line( 1, 5){30}}
\put( 60,120) {\line( 1,-3){30}}
\put( 90,140) {\circle*{5}}
\thinlines
\put( 69,126) {\vector(-3,-2){6}}
\put( 84,136) {\vector( 3, 2){4}}
\multiput( 90,140)(-3,-2){10}{\circle*{1}}
\multiput( 90,140)( 4.5,0.75){13}{\circle*{1}}
\multiput( 90,140)(-4.5,0.75){13}{\circle*{1}}
\multiput( 90,140)(0,5){26}{\circle*{1}}
\put( 90,270) {\vector( 0, 1){40}}
\multiput( 90,140)( 4, 0){8}{\circle*{1}}
\put(120,140) {\vector( 1, 0){90}}
\put( 60,120) {\vector(-3,-2){40}}
\multiput( 90, 30)( 5, 0){24}{\line(1,0){2}}
\multiput( 90,270)( 5, 0){24}{\line(1,0){2}}
\put(190, 30) {\vector( 0, 1){110}}
\put(190,140) {\vector( 0, 1){130}}
\put(190,270) {\vector( 0,-1){130}}
\put(190,140) {\vector( 0,-1){110}}
\put(200, 80) {$\beta_2$}
\put(200,200) {$\beta_2$}
\put( 65,133) {$\beta_1$}
\put(220,135) {$\hat y$}
\put( 88,320) {$\hat z$}
\put( 10, 83) {$\hat x$}
\end{picture}
\caption{A bi-pyramidal structure with ${\cal D}_{3h}$ symmetry.}
\label{bipyramid}
\end{figure}
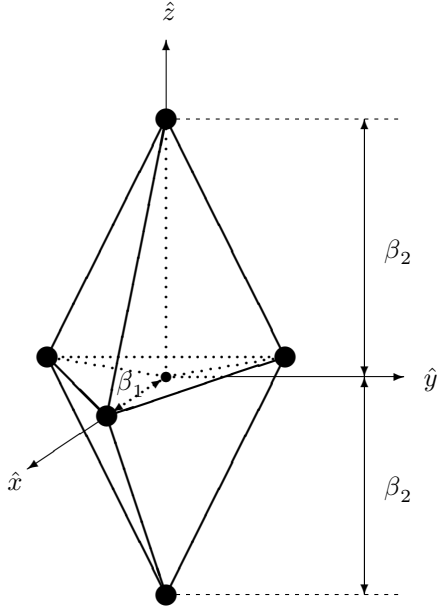

The starting point of our study is the cluster structure of $^{20}$Ne. This
structure has been very recently analyzed \cite{bijker5}. Ground state
properties of $^{20}$Ne appear to be well described by a bi-pyramidal
configuration, $\alpha$-$^{12}$C-$\alpha$, with $D_{3h}$ symmetry,
Fig.~\ref{bipyramid}. The coordinates ($r_{i},\theta_{i},\varphi_{i}$) 
of the five constituent $\alpha $-particles are 
\begin{eqnarray}
(r_{1},\theta_{1},\varphi_{1}) &=& (\beta_{1},\tfrac{\pi}{2},0) ~,
\nonumber\\
(r_{2},\theta_{2},\varphi_{2}) &=& (\beta_{1},\tfrac{\pi}{2}, 
\tfrac{2\pi}{3}) ~,
\nonumber\\
(r_{3},\theta_{3},\varphi_{3}) &=& (\beta_{1},\tfrac{\pi}{2},
\tfrac{4\pi}{3}) ~,
\nonumber\\
(r_{4},\theta_{4},\varphi_{4}) &=& (\beta_{2},0,-) ~,
\nonumber\\
(r_{5},\theta_{5},\varphi_{5}) &=& (\beta_{2},\pi,-) ~.
\end{eqnarray}%
Assuming a gaussian distribution for the density of the $\alpha $-particles,%
\begin{equation}
\rho _{\alpha }(\vec{r})=\left( \frac{\alpha }{\pi }\right) ^{3/2}e^{-\alpha
r^{2}} ~,
\end{equation}%
we have an overall density%
\begin{eqnarray}
\rho (\vec{r}) &=&\frac{1}{5}\left( \frac{\alpha _{1}}{\pi }\right)
^{3/2}\sum_{i=1}^{3}\exp \left[ -\alpha _{1}\left( \vec{r}-\vec{r}%
_{i}\right) ^{2}\right]  \nonumber \\
&&+\frac{1}{5}\left( \frac{\alpha _{2}}{\pi }\right)
^{3/2}\sum_{i=4}^{5}\exp \left[ -\alpha _{2}\left( \vec{r}-\vec{r}%
_{i}\right) ^{2}\right] ~.
\label{density}
\end{eqnarray}%
In the calculation of \cite{bijker5}, it was assumed that all $\alpha $%
-particles are identical with $\alpha _{1}=\alpha _{2}=\alpha =0.53$ fm$^{-2}
$ obtained from the r.m.s. radius $\left\langle r^{2}\right\rangle _{\alpha
}^{1/2}=1.674(12)$ fm of the free $\alpha $-particle. All properties of the
ground state of $^{20}$Ne can then be determined in terms of two parameters $%
\beta_{1}$ and $\beta_{2}$. A fit to the elastic form factor in electron
scattering gives $\beta_{1}=1.82$ fm, $\beta _{2}=3.00$ fm, from which one
calculates the r.m.s. radius $\left\langle r^{2}\right\rangle ^{1/2}=2.89$
fm, the intrinisc quadrupole moment $Q_{0}=52.5$ efm$^{2}$ and the quenched
moment of inertia $I/m=97.6$ fm$^{2}$ \cite{bijker5}.
We note incidentally here that while the density Eq.~(\ref{density}) describes well the r.m.s. radius and intrinsic quadrupole moment, it does not describe well the moment of inertia and a quenched value had to be used in \cite{bijker5}. This may be due to the neglect of Pauli exchange contributions which make the cluster structure tighter than that given in Eq.~(\ref{density}) (see the comparison between quenched and unquenched values in Section~3.4 of \cite{bijker5}).

The excitation spectrum of $^{20}$Ne consists of the vibrational states of
the bi-pyramidal configuration \cite{bijker5}. There are $15-6=9$ vibrations,
three singly degenerate and three doubly degenerate. The species of these
vibrations, their characterization and numbering are given in \cite{bijker5}. 
In characterizing the representations we can use the representations of
the group $D_{3h}$ which is the symmetry group of the bi-pyramid of Fig.~\ref{bipyramid}. 
The energy levels of $^{20}$Ne can then be written as those of a symmetric
top (rotational part) and in the harmonic approximation (vibrational part) as 
\begin{equation}
E\left( \left[ v\right] ,K,L\right) =E_{0}+B_{x\left[ v\right] }L(L+1)+\left[
B_{z}-B_{x}\right] _{\left[ v\right] }K^{2}+\sum_{i=1}^{6}\omega _{i}v_{i} ~.
\end{equation}%
where $\left[ v\right] \equiv \left[ v_{1},v_{2},v_{3},v_{4},v_{5},v_{6}%
\right] $, $L$ is the angular momentum and $K$ the projection on the
intrinsic axis ($z$-axis in Fig.~\ref{bipyramid}). The vibrational and rotational
parameters of $^{20}$Ne are summarized in Table~\ref{gsvib}. Here $\Gamma $ denotes
the representations of ${\cal D}_{3h}$, $K$ the projection of the angular momentum
on the $z$-axis and $P$ the parity.

\begin{table}
\centering
\caption[Rotational and vibrational parameters]
{Summary of rotational and vibrational parameters in $^{20}$Ne \cite{bijker5}.}
\label{gsvib}
\vspace{10pt}
\begin{tabular}{ccccc}
\hline
\noalign{\smallskip}
& $\Gamma$ & $K^P$ & $\omega$ (MeV) & $B$ (keV) \\
\noalign{\smallskip}
\hline
\noalign{\smallskip}
g.s.    & $A'_1$  & $0^+$ & 0.00 & 212 \\ 
$v_{1}$ & $A''_2$ & $0^-$ & 5.52 & 137 \\
$v_{2}$ & $A'_1$  & $0^+$ & 6.72 & 127 \\
$v_{3}$ & $A'_1$  & $0^+$ & 7.19 & 130 \\ 
$v_{4}$ & $E'$    & $1^-$ & 8.59 & 134 \\
        &         & $2^+$ & 8.53 & 112 \\
$v_{5}$ & $E'$    & $1^-$ & 8.42 & 147 \\ 
        &         & $2^+$ & 8.71 & 130 \\ 
$v_{6}$ & $E''$   & $1^+$ & 9.68 & 124 \\
        &         & $2^-$ & 4.10 & 145 \\ 
\noalign{\smallskip}
\hline
\end{tabular}
\end{table}

In the following sections, we make also reference to the cluster structure
of $^{19}$Ne and $^{19}$F. These nuclei can be viewed either as a hole in
the structure of $^{20}$Ne, or equivalently as a bi-pyramid in which one 
$\alpha$ particle on the $z$-axis of Fig.~\ref{bipyramid} is replaced by $^{3}$H (in $^{19}$F) or $^{3}$He (in $^{19}$Ne), in which case the symmetry of the cluster is
reduced from $D_{3h}$ to $D_{3}$, with composition $\alpha$ - $^{12}$C - $^{3}$H 
or $\alpha$ - $^{12}$C - $^{3}$He.

\section{Structure of $^{21}$Ne}
\label{sec3}

Extensive sets of experimental data have been accumulated in the last 50
years. We use here the recent compilation of Firestone \cite{firestone}. In
order to compare with theoretical models we first make assingments of the
observed energy levels into rotational bands characterized by a value of $J$
and $K$. The rotational bands are analyzed with the formula%
\begin{eqnarray}
E_{rot}(\Omega,K,J^P) &=& \varepsilon_{\Omega} 
+ B_{\Omega} \left[ J(J+1) - b_{\Omega} K^{2} \right.
\nonumber\\
&& \left. + a_{\Omega}(-1)^{J+1/2} \left( J+\tfrac{1}{2} \right) 
\delta_{K,1/2} \right] 
\label{erot}
\end{eqnarray}
where $\Omega$ labels the rotational bands in $^{21}$Ne, 
$\varepsilon_{\Omega }$ is the intrinsic energy, $B_{\Omega}$ is the inertial parameter $B_{\Omega}=\hbar^{2}/2I$, $a_{\Omega }$ the decoupling parameter and $b_{\Omega}$ has contributions from both the symmetric top \cite{bijker5}
and the Coriolis term \cite{preston}
\begin{equation}
b_{\Omega}=\frac{B_{x\left[ v\right] }-B_{z\left[ v\right] }}{B_{x\left[ v%
\right] }}+2\cong 1,
\end{equation}
since the first term is $\sim -1$ in $^{20}$Ne as given in Eq.~(30) of \cite%
{bijker5}. In the following analysis we take $b_{\Omega }=1$.

\subsection{Assignments of states to bands}

\begin{figure}
\includegraphics[scale=0.8]{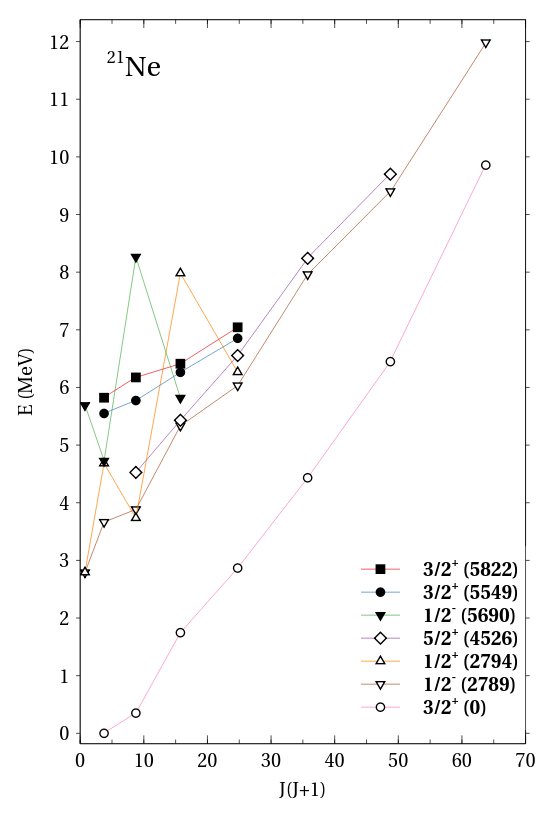}
\caption[Rotational bands in $^{21}$Ne]
{Energies of assigned states in $^{21}$Ne to rotational bands as a 
function of $J(J+1)$.}
\label{bandsNe}
\end{figure}

We identify bands by the value of $K^{P}$ and the energy of the $J^{P}=K^{P}$
state in the band. States are assigned on the basis of their energy, their
electromagnetic transition rates and their branching ratios when available,
and of their spectroscopic factors in $(d,p)$ and $(p,d)$ reactions. For the
low-lying states the assignments are straightforward, but at higher
excitation energy they are rather difficult since there are several states
with the same spin and parity. We have been able to identify seven bands as
shown in Table~\ref{statesNe} and Fig.~\ref{bandsNe}. The superscript $^{a}$ in 
Table~\ref{statesNe} denotes states with no spin assignment in \cite{wheldon}.

\begin{table}
\centering
\caption{Rotational bands in $^{21}$Ne}
\label{statesNe}
\vspace{10pt}
\begin{tabular}{ccrr}
\hline
\noalign{\smallskip}
$K^{P}(E_{\rm exc})$ & $J^{P}$ & $E_{\rm exp}$ & $E_{\rm th}$ \\ 
\noalign{\smallskip}
\hline
\noalign{\smallskip}
$3/2^{+}$(0) & $3/2^{+}$ & 0 & 0 \\ 
&   $5/2^{+}$  &  351 &  690 \\ 
&   $7/2^{+}$  & 1746 & 1656 \\ 
&   $9/2^{+}$  & 2867 & 2898 \\ 
&  $11/2^{+}$  & 4433 & 4416 \\ 
& $(13/2^{+})$ & 6448 & 6120 \\ 
& $(15/2^{+})$ & 9857 & 8280 \\
\noalign{\smallskip}
$1/2^{-}$(2789) & $1/2^{-}$ & 2789 & 2789 \\ 
&   $3/2^{-}$  &  3664 &  3473 \\ 
&   $5/2^{-}$  &  3884 &  3813 \\ 
&   $7/2^{-}$  &  5334 &  5409 \\ 
&   $9/2^{-}$  &  6033 &  6021 \\ 
& $(11/2^{-})$ &  7961 &  8529 \\ 
& $(13/2^{-})$ &  9401 &  9413 \\ 
& $(15/2^{-})$ & 11984 & 12834 \\ 
\noalign{\smallskip}
$1/2^{+}$(2794) & $1/2^{+}$ & 2794 & 2794 \\ 
& $3/2^{+}$ & 4684 & 4675 \\ 
& $5/2^{+}$ & 3736 & 3680 \\ 
& $7/2^{+}$ & 7982 & 8069 \\ 
& $9/2^{+}$ & 6267 & 6278 \\ 
\noalign{\smallskip}
$5/2^{+}$(4526) & $5/2^{+}$ & 4526 & 4526 \\ 
&  $(7/2^{+})$ & 5431 & 5443 \\ 
&   $9/2^{+}$  & 6554 & 6622 \\ 
& $(11/2^{+})$ & 8240 & 8063 \\ 
& $(13/2^{+})$ & 9700 & 9766 \\
\noalign{\smallskip}
$1/2^{-}$(5690) & $1/2^{-}$ & 5690 & 5690 \\ 
&   $3/2^{-}$  &  4725        & 4725 \\   
&  $(5/2^{-})$ &  8264        & 8071 \\ 
&   $7/2^{-}$  &  5818        & 5820 \\ 
\noalign{\smallskip}
$3/2^{+}$(5549) & $3/2^{+}$ & 5549 & 5549 \\ 
& $(5/2^{+})$ & 5773 & 5849 \\ 
& $(7/2^{+})$ & 6263 & 6269 \\ 
&  $9/2^{+}$  & 6853 & 6809 \\ 
\noalign{\smallskip}
$3/2^{+}$(5822) & $3/2^{+}$ & 5822 & 5822 \\ 
& $(5/2^{+})$ &  6174        & 6112 \\ 
& $(7/2^{+})$ & $(6412)^{a}$ & 6518 \\ 
& $(9/2^{+})$ &  7044        & 7040 \\
\noalign{\smallskip}
\hline
\end{tabular}
\end{table}

\section{Cluster interpretation of rotational bands}
\label{sec4}

In order to classify the single particle intrinsic states, $\Omega $, we use
the cluster shell model (CSM) \cite{dellarocca1}. 

\subsection{Cluster shell model}

In the CSM, intrinsic states in $^{21}$Ne are calculated by solving the Schr\"{o}dinger equation 
\begin{equation}
H=\frac{\vec{p}^{2}}{2m}+V(\vec{r})+V_{so}(\vec{r}) ~,
\label{hcsm}
\end{equation}
in the potential generated by the appropriate geometric confguration. For
the bi-pyamidal structure of Fig.~\ref{bipyramid}, the potential is given by \cite%
{dellarocca1} 
\begin{eqnarray}
V(\vec{r}) &=& -V_{1} \sum_{\lambda \mu} 
f_{\lambda}^{(1)}(r) Y_{\lambda \mu}(\theta,\phi)
\sum_{i=1}^{3} Y_{\lambda \mu}^{\ast}(\theta_{i},\phi_{i}) 
\nonumber \\
&& -V_{2} \sum_{\lambda \mu} 
f_{\lambda}^{(2)}(r) Y_{\lambda \mu}(\theta,\phi)
\sum_{i=4}^{5} Y_{\lambda \mu}^{\ast}(\theta_{i},\phi_{i}) ~,
\label{vr}
\end{eqnarray}%
where%
\begin{eqnarray}
f_{\lambda}^{(1)}(r) &=& e^{-\alpha_{1}(r^{2}+\beta_1^2)} \, 4\pi \, 
i_{\lambda}(2\alpha_{1}\beta_1 r) ~,
\nonumber\\
f_{\lambda}^{(2)}(r) &=& e^{-\alpha_{2}(r^{2}+\beta_2^2)} \, 4\pi \, 
i_{\lambda}(2\alpha_{2}\beta_2 r) ~, 
\label{fl}
\end{eqnarray}
and the spin-orbit potential by
\begin{eqnarray}
V_{\rm so}(\vec{r}) &=& V_{1,{\rm so}} \sum_{\lambda \mu} \left(  
- 2\alpha_{1} f_{\lambda}^{(1)}(r) + \frac{\lambda}{r^{2}}f_{\lambda}^{(1)}(r) 
+ \frac{2\alpha_{1}\beta_1}{r} f_{\lambda+1}^{(1)}(r) \right) 
\nonumber\\
&& \quad \times \frac{1}{2} 
\left[ Y_{\lambda \mu}(\theta,\phi) (\vec{s} \cdot \vec{l}) 
+ (\vec{s} \cdot \vec{l}) Y_{\lambda \mu}(\theta,\phi )\right]
\sum_{i=1}^{3} Y_{\lambda \mu}^{\ast}(\theta_{i},\phi_{i})  
\nonumber \\
&& + V_{2,{\rm so}} \sum_{\lambda \mu} \left(  
- 2\alpha_{2} f_{\lambda}^{(2)}(r) + \frac{\lambda}{r^{2}}f_{\lambda}^{(2)}(r) 
+ \frac{2\alpha_{2}\beta_2}{r} f_{\lambda+1}^{(2)}(r) \right) 
\nonumber\\
&& \quad \times \frac{1}{2} 
\left[ Y_{\lambda \mu}(\theta,\phi) (\vec{s} \cdot \vec{l}) 
+ (\vec{s} \cdot \vec{l}) Y_{\lambda \mu}(\theta,\phi )\right]
\sum_{i=4}^{5} Y_{\lambda \mu}^{\ast}(\theta_{i},\phi_{i}) ~.
\label{vso}
\end{eqnarray}
The values of $\alpha_{1}$ and $\alpha_{2}$ that appear in $f_{\lambda}$
are not the same as those appearing in the density, since they are obtained
by convoluting the density with the $\alpha$-nucleon interaction as
discussed in Eq.~(13) of \cite{dellarocca1}. In both equations we have
allowed for the possibility of a different interaction of the nucleon with
the $\alpha $-particles in the $xy$-plane and with those along the $z$ axis 
($\alpha_{1} \neq \alpha_{2}$).

\begin{figure}
\centering
\setlength{\unitlength}{0.95pt}
\begin{picture}(340,550)(10,10)
\thicklines
\put ( 30, 10) {\line(0,1){550}}
\put ( 30, 20) {\line(1,0){2}}
\put ( 30, 40) {\line(1,0){2}}
\put ( 30, 60) {\line(1,0){2}}
\put ( 30, 80) {\line(1,0){2}}
\put ( 28,100) {\line(1,0){4}}
\put ( 30,120) {\line(1,0){2}}
\put ( 30,140) {\line(1,0){2}}
\put ( 30,160) {\line(1,0){2}}
\put ( 30,180) {\line(1,0){2}}
\put ( 28,200) {\line(1,0){4}}
\put ( 30,220) {\line(1,0){2}}
\put ( 30,240) {\line(1,0){2}}
\put ( 30,260) {\line(1,0){2}}
\put ( 30,280) {\line(1,0){2}}
\put ( 28,300) {\line(1,0){4}}
\put ( 30,320) {\line(1,0){2}}
\put ( 30,340) {\line(1,0){2}}
\put ( 30,360) {\line(1,0){2}}
\put ( 30,380) {\line(1,0){2}}
\put ( 28,400) {\line(1,0){4}}
\put ( 30,420) {\line(1,0){2}}
\put ( 30,440) {\line(1,0){2}}
\put ( 30,460) {\line(1,0){2}}
\put ( 30,480) {\line(1,0){2}}
\put ( 28,500) {\line(1,0){4}}
\put ( 30,520) {\line(1,0){2}}
\put ( 30,540) {\line(1,0){2}}
\put ( 12, 97) {-20}
\put ( 12,197) {-15}
\put ( 12,297) {-10}
\put ( 16,397) {-5}
\put ( 20,497) {0}
\put (  5,435) {\rotatebox{90}{E(MeV)}}
\put( 65, 35.8) {\line(1,0){15}}
\put( 37, 33.8) {-23.21}
\put( 82, 33.8) {$\frac{1}{2}^+$}
\put( 95, 33.8) {-15.99}
\put(125, 33.8) {+1.01}
\put( 65,166.0) {\line(1,0){15}}
\put( 37,163.0) {-16.70}
\put( 82,163.0) {$\frac{1}{2}^-$}
\put( 95,163.0) {-9.48}
\put(125,163.0) {-1.64}
\put( 65,235.8) {\line(1,0){15}}
\put( 37,232.8) {-13.21}
\put( 82,232.8) {$\frac{3}{2}^-$}
\put( 95,232.8) {-5.99}
\put( 65,291.0) {\line(1,0){15}}
\put( 37,287.0) {-10.45}
\put( 82,287.0) {$\frac{1}{2}^-$}
\put( 95,287.0) { -3.23}
\put(125,287.0) {+0.64}
\put( 65,303.2) {\line(1,0){15}}
\put( 37,301.2) { -9.84}
\put( 82,301.2) {$\frac{1}{2}^+$}
\put( 95,301.2) { -2.62}
\put(125,301.2) {+2.16}
\put( 65,355.6) {\line(1,0){15}}
\put( 37,352.6) { -7.22}
\put( 82,352.6) {$\frac{3}{2}^+$}
\put( 95,352.6) {  0}
\put( 65,391.2) {\line(1,0){15}}
\put( 37,388.2) { -5.55}
\put( 82,388.2) {$\frac{1}{2}^+$}
\put( 95,388.2) { 1.78}
\put(125,388.2) {+0.95}
\put( 65,412.6) {\line(1,0){15}}
\put( 37,409.6) { -4.37}
\put( 82,409.6) {$\frac{5}{2}^+$}
\put( 95,409.6) { 2.85}
\put( 65,442.4) {\line(1,0){15}}
\put( 37,434.2) { -2.88}
\put( 82,434.2) {$\frac{1}{2}^+$}
\put( 95,434.2) { 4.34}
\put(125,434.2) {-1.11}
\put( 65,445.2) {\line(1,0){15}}
\put( 37,447.2) { -2.74}
\put( 82,447.2) {$\frac{1}{2}^-$}
\put( 95,447.2) { 4.48}
\put(125,447.2) {-2.60}
\put( 65,487.0) {\line(1,0){15}}
\put( 37,480.0) { -0.65}
\put( 82,480.0) {$\frac{3}{2}^-$}
\put( 95,480.0) { 6.57}
\put( 65,492.6) {\line(1,0){15}}
\put( 37,493.6) { -0.37}
\put( 82,493.6) {$\frac{3}{2}^+$}
\put( 95,493.6) { 6.85}
\put( 40,520.0) {$\quad \varepsilon'_K$}
\put( 82,520.0) {$K^P$}
\put( 95,520.0) {$\quad \varepsilon_K$}
\put(125,520.0) {$\quad a$}
\put( 82,550.0) {CSM}
\put(190,294.4) {\line(1,0){15}}
\put(162,287.4) {-10.28}
\put(207,287.4) {$\frac{1}{2}^+$}
\put(220,287.4) { -3.30}
\put(250,287.4) {+1.86}
\put(190,300.8) {\line(1,0){15}}
\put(162,301.8) { -9.96}
\put(207,301.8) {$\frac{1}{2}^-$}
\put(220,301.8) { -3.04}
\put(250,301.8) {+0.54}
\put(190,360.6) {\line(1,0){15}}
\put(162,357.6) { -6.97}
\put(207,357.6) {$\frac{3}{2}^+$}
\put(220,357.6) {  0}
\put(190,448.8) {\line(1,0){15}}
\put(162,445.8) { -2.56}
\put(207,445.8) {$\frac{5}{2}^+$}
\put(220,445.8) { 4.41}
\put(190,467.0) {\line(1,0){15}}
\put(162,464.0) { -1.65}
\put(207,464.0) {$\frac{1}{2}^-$}
\put(220,464.0) { 5.32}
\put(250,464.0) {-2.85}
\put(165,520.0) {$\quad \varepsilon'_K$}
\put(207,520.0) {$K^P$}
\put(220,520.0) {$\quad \varepsilon_K$}
\put(250,520.0) {$\quad a$}
\put(207,550.0) {Exp}
\put(290,264.4) {\line(1,0){15}}
\put(307,261.4) {$\frac{1}{2}^-$}
\put(320,261.4) { -4.56}
\put(290,278.6) {\line(1,0){15}}
\put(307,275.6) {$\frac{1}{2}^+$}
\put(320,275.6) { -3.85}
\put(290,355.6) {\line(1,0){15}}
\put(307,352.6) {$\frac{3}{2}^+$}
\put(320,352.6) {  0}
\put(290,412.0) {\line(1,0){15}}
\put(307,409.0) {$\frac{1}{2}^+$}
\put(320,409.0) { 2.82}
\put(290,441.0) {\line(1,0){15}}
\put(307,432.0) {$\frac{5}{2}^+$}
\put(320,432.0) { 4.27}
\put(290,442.8) {\line(1,0){15}}
\put(307,445.8) {$\frac{1}{2}^-$}
\put(320,445.8) { 4.36}
\put(307,520.0) {$K^P$}
\put(320,520.0) {$\quad \varepsilon_K$}
\put(300,550.0) {Nilsson}
\put(245,100.0) {\Large $^{21}$Ne}
\end{picture}
\caption[Intrinsic energies in $^{21}$Ne]{Energies of the intrinsic neutron states in $^{21}$Ne in the CSM (left) compared with the experimental intrinsic energies (middle) and the Nilsson model (right) \cite{nilsson}.}
\label{defspNe}
\end{figure}
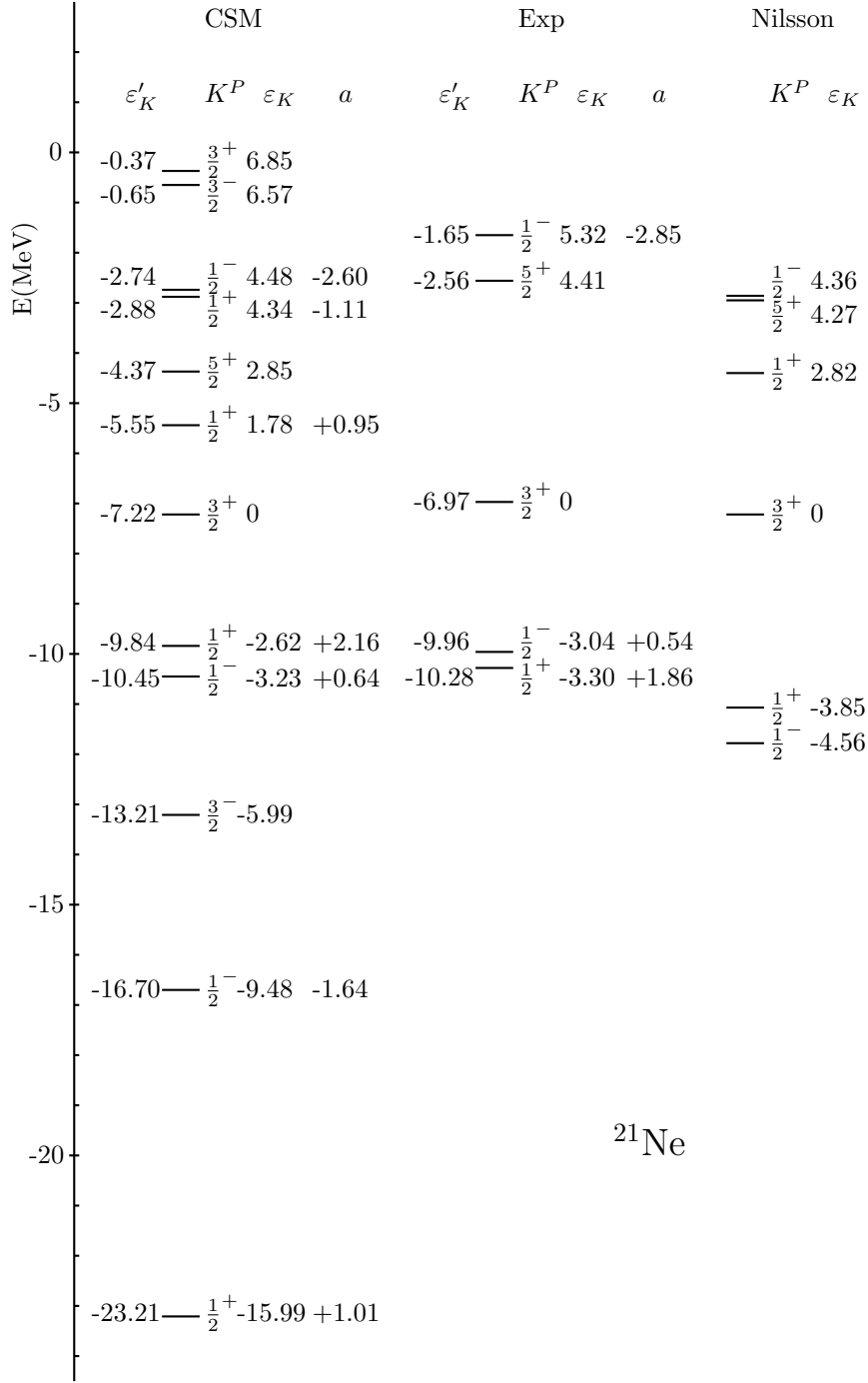

The intrinsic states in the potential of Eqs.~(\ref{vr}-\ref{vso}) are characterized by
the value of the projection $K$ of the angular momentum of the single
particle on the intrinsic $z$ axis, $\left\langle j_{3}\right\rangle $, and
the parity $P$. A calculation of the single-particle levels in a potential
with parameters $V_{1}=5.4$ MeV, $V_{1,\rm so}=5.4$ MeVfm$^{2}$, $\alpha
_{1}=0.0679$ fm$^{-2}$, $V_{2}=24.5$ MeV, $V_{2,\rm so}=24.5$ MeVfm$^{2}$, $%
\alpha_{2}=0.0679$ fm$^{-2}$ at the values of the deformation parameters $%
\beta_{1}=1.80$ fm, $\beta _{2}=3.00$ fm of $^{20}$Ne \cite{bijker5} gives the
results in the left column of Fig.~\ref{defspNe}. The calculation has been carried out in the $Z_2$ basis using a model space with 10 oscillator shells, $\nu =0.1833$ fm$^{-2}$ and $\hslash \omega=15.2$ MeV. In this figure, both the intrinsic energies from the $^{20}$Ne+n threshold, denoted by $\varepsilon_{K}^{\prime}$ in this article, and
those relative to the energy of the $K^{P}=3/2^{+}$ intrinsic state at $%
\varepsilon_{K}^{\prime}=-7.22$ MeV, denoted by $\varepsilon_{K}$ in this
article, are given, as well as the decoupling parameters $a_{\Omega}$ 
calculated accordingly to 
\begin{equation}
a_{\Omega}=-\sum_{nlj} (-1)^{j+1/2} \left( j+\tfrac{1}{2}\right)
\left\vert c_{nlj,1/2}^{\Omega} \right\vert^{2} ~,
\end{equation}%
where $c_{nljm}^{\Omega}$ are the expansion coefficients of the intrinsic wave
functions in the spherical basis%
\begin{equation}
\left\vert \chi_{\Omega} \right\rangle
=\sum_{nljm}c_{nljm}^{\Omega}\left\vert nljm \right\rangle ~.
\end{equation}%
The decoupling parameter only contributes to $K=1/2$ bands. 
The relation between $\varepsilon_{K}$ and $\varepsilon_{K}^{\prime}$ is 
\begin{equation}
\varepsilon_{K}=\varepsilon_{K}^{\prime}-\varepsilon_{K^{P}=3/2^{+}}^{\prime} ~.
\end{equation}

In order to compare the calculated values with the experimental values and
thus to provide an intepretation of the observed bands, one can first use
Eq.~(\ref{erot}) to extract the values of $\varepsilon ,B,$ and $a$ for each of the
observed bands of Sect.~\ref{sec3}. From the value of $\varepsilon $ one can then
obtain the values of $\varepsilon_{K}$. Since we shall consider both
particle (p) and hole (h) states, the relatioship between $\varepsilon_{K}$
and $\varepsilon$ is
\begin{eqnarray}
p &:&\varepsilon _{K}=\varepsilon -\varepsilon _{K^{P}=3/2^{+}} ~,  
\nonumber\\
h &:&\varepsilon _{K}=-\varepsilon -\varepsilon _{K^{P}=3/2^{+}} ~.
\end{eqnarray}%
From the values of $\varepsilon _{K}$ we can then obtain the values of 
$\varepsilon_{K}^{\prime}$ using
\begin{equation}
\varepsilon _{K}^{\prime }=\varepsilon _{K}-S_{n}-\frac{3}{2}B_{K^{P}=3/2^{+}} ~,
\end{equation}%
where $S_{n}$ is the neutron separation energy $S_{n}=6761.16(4)$ keV and we
have subtracted the energy of the ground state $J^P=K^P=3/2^{+}$. The experimental values of the particle and hole intrinsic
states $\varepsilon_{K}(\exp)$, $\varepsilon_{K}^{\prime}(\exp)$ and
the experimental decoupling parameters, $a(\exp)$ are shown in the center
column of Fig.~\ref{defspNe}.

\subsection{Assignments into bands}

\subsubsection{$K^{P}=3/2^{+}(0)$}

This band is the ground state band. It is observed up to $J^{P}=15/2^{+}$
with inertial parameter $B=138$ keV and intrinsic energy relative to the
ground state $J^{P}=3/2^{+}$ of $\varepsilon =-0.207$ MeV. This band has an
anomaly at $J^{P}=5/2^{+}$ since the energy of this state does not fit the
rotational behavior, Eq.~(\ref{erot}). This band with $\varepsilon_{K}^{\prime
}=-6.97 $ MeV relative to the $^{20}$Ne+n threshold can be associated with
the calculated $K^{P}=3/2^{+}$ band at $\varepsilon _{K}^{\prime }=-7.22$
MeV.

\subsubsection{$K^{P}=1/2^{-}(2789)$}

This band starts at $E_{x}=2789$ keV. It is observed up to $J^{P}=15/2^{-}$
with inertial parameter $B=148$ keV, decoupling parameter $a=0.54$ and
intrinsic energy $\varepsilon =2.824$ MeV. This band can be associated with
the calculated $K^{P}=1/2^{-}$ band at $\varepsilon _{K}^{\prime }=-10.45$
MeV. The interpretation of this band as a "hole" band was suggested years
ago by Howard \textit{et al.} \cite{howard1} on the basis of intensities in
$(d,p)$ and $(p,d)$ reactions, following a previous theoretical article \cite{howard2} and confirmed by further experiments \cite{stanford}. The observed
value of $\varepsilon_{K}^{\prime}=-10.0$ MeV in this interpretation is in
excellent agreement with the calculated value, as well as with the
calculated value of the decoupling parameter $a=0.64$. Further confirmation
of this band as a "hole" band is obtained from an analysis of a similar band
in $^{19}$F and $^{19}$Ne.

\subsubsection{$K^{P}=1/2^{+}(2794)$}

This band starts at $E_{x}=2794$ keV. It is observed up to $J^{P}=9/2^{+}$
with inertial parameter $B=214$ keV, decoupling parameter $a=1.93$ and
intrinsic energy $\varepsilon =3.088$ MeV. This band can be associated with
the calculated $K^{P}=1/2^{+}$ band at $\varepsilon _{K}^{\prime}=-9.84$
MeV. The interpretation of this band as a "hole" band was also suggested in 
\cite{howard1}. The observed value $\varepsilon_{K}^{\prime}=-10.3$ MeV is
in excellent agreement with the calculated value, as it is the decoupling
parameter calculated at $a=2.16$. Further confirmation is obtained from an
analysis of a similar band in $^{19}$F, $^{19}$Ne. The almost equal energy
of the $1/2^{+}$ state of this band with the $1/2^{-}$ state of the band $%
K^{P}=1/2^{-}(2789)$ suggests that these two bands form a parity doublet.

\subsubsection{$K^{P}=5/2^{+}(4526)$}

This band starts at $E_{x}=4526$ keV. It is observed up to $K^{P}=13/2^{+}$
with inertial parameter $B=128$ keV, and intrinsic energy $\varepsilon
=4.199 $ MeV. This band can be associated with the calculated $K^{P}=5/2^{+}$
"particle" band at $\varepsilon _{K}^{\prime }=-4.37$ MeV. The observed
value $\varepsilon _{K}^{\prime }=-2.56$ MeV is however only in qualitative
agreement with the calculated value. Howard \textit{et al.} \cite{howard1}
also associate this band with the $K^{P}=5/2^{+}$ "particle" band.

\subsubsection{$K^{P}=1/2^{-}(5690)$}

This band poses some difficulties as the lowest member of the band has $%
J^{P}=3/2^{-}$ and it has been the subject of many different
interpretations. Our analysis agrees with the assignment of \cite{howard1} as
a "particle" intrinsic state originating from the $1f_{7/2}$ spherical
state. The extracted values of $\varepsilon =5.107$ MeV, ($\varepsilon
_{K}^{\prime }=-1.65$ MeV) and $a=-2.85$ are in agreement with the
calculated values $\varepsilon _{K}^{\prime }=-2.88$ MeV, $a=-2.60$. The
large and negative value of $a$ accounts for the inversion in the lowest
member of the band. The inertial parameter of this band is $B=174$ MeV. This
interpretation is strongly supported by $^{20}$Ne$(d,p)$ data \cite{howard1}.
It is in disagreement with the interpretation of this band as the parity
doubled band of the ground state.

\subsubsection{$K^{P}=3/2^{+}(5549)$}

This band starts at $E_{x}=5549$ keV. It is observed up to $J^{P}=9/2^{+}$
with a small inertial parameter $B=60$ keV and $\varepsilon =5.459$ MeV. One
possible interpretation of this band is as the single "particle" state with $%
K^{P}=3/2^{+}$ calculated at $\varepsilon _{K}^{\prime }=-0.37$ MeV. Another
possible interpretation is that of a single particle state $K^{P}=3/2^{+}$
at $\varepsilon _{K}^{\prime }=-7.22$ MeV coupled to the vibrational $%
A_{1}^{\prime }:K^{P}=0^{+}(6.72)$ of $^{20}$Ne. The small inertial
parameter $B=60$ keV and the calculated value $\varepsilon _{K}^{\prime
}=-0.50$ MeV support this interpretation, which we the tentatively adopt.

\subsubsection{$K^{P}=3/2^{+}(5826)$}

The interpretation of this band poses several difficulties. If it assumed to
have $K^{P}=3/2^{+}$ this band starts at $E_{x}=5826$ keV. It is observed up
to $J^{P}=9/2^{+}$ with a small inertial parameter $B=58$ keV and $%
\varepsilon =5.735$ MeV, ($\varepsilon _{K}^{\prime }=-1.02$ MeV). Its
interpretation is a single particle state with $K^{P}=3/2^{+}$ at $%
\varepsilon _{K}^{\prime }=-7.22$ MeV coupled to the vibrational state $%
A_{1}^{\prime }:K^{P}=0^{+}(7.19)$ of $^{20}$Ne. The small inertial
parameter $B=58$ keV and the calculated value $\varepsilon _{K}^{\prime
}=-0.03$ MeV support this intepretation. Also note that the difference $%
\Delta \varepsilon _{K}=-0.28$ MeV between this band and the band $%
K^{P}=3/2^{+}(5549)$ is in agreement with the difference $\Delta \varepsilon
_{K}=-0.47$ MeV in $^{20}$Ne. An alternative interpretation is that,
together with a state $J^{P}=1/2^{+}$ at $5525$ keV, it forms a $%
K^{P}=1/2^{+}$ rotational band. This interpretation is however disfavored by
the small inertial parameter and by the lack of spin asignment for the $5525$
keV level.

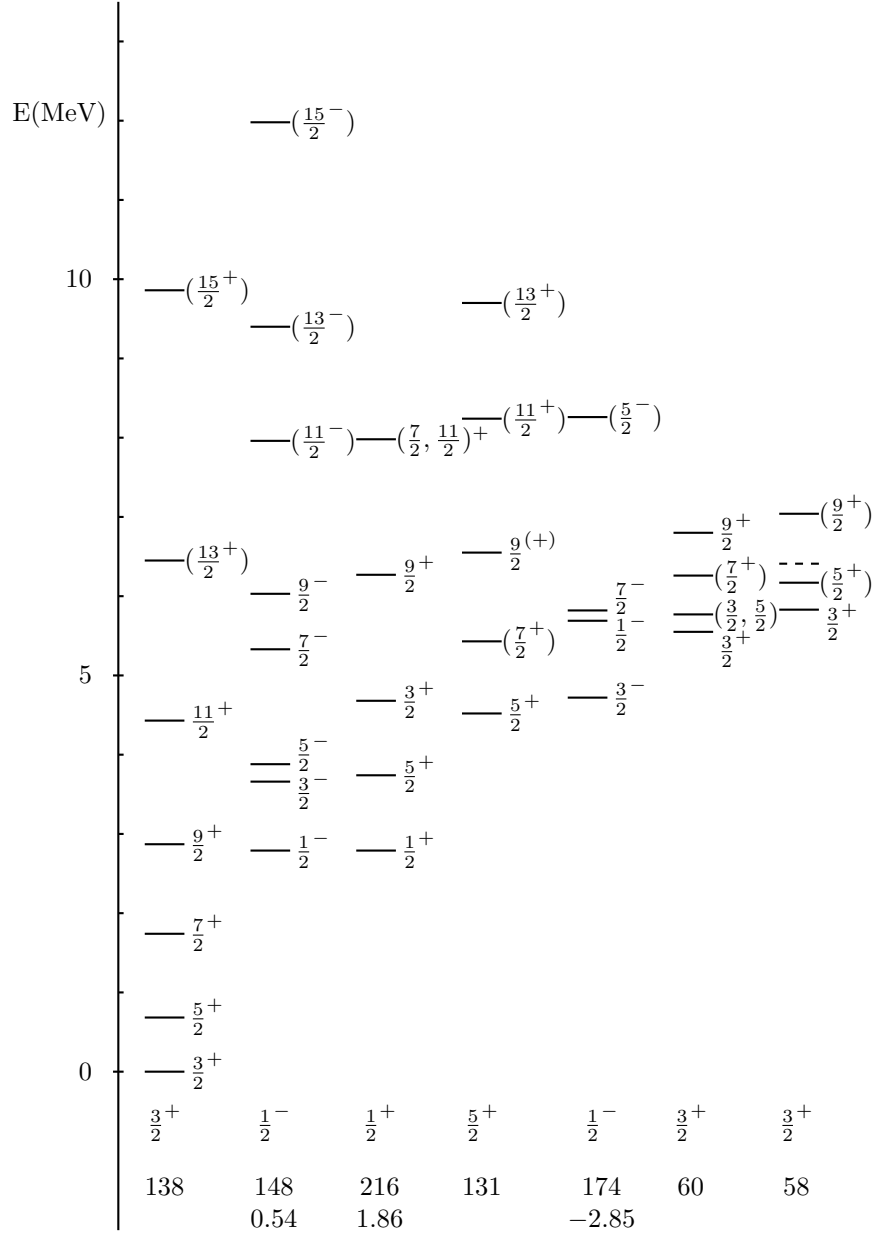
\begin{figure}
\centering
\setlength{\unitlength}{1pt}
\begin{picture}(340,465)(-10,-30)
\thicklines
\put ( 30,-30) {\line(0,1){465}}
\put ( 28, 30) {\line(1,0){4}}
\put ( 30, 60) {\line(1,0){2}}
\put ( 30, 90) {\line(1,0){2}}
\put ( 30,120) {\line(1,0){2}}
\put ( 30,150) {\line(1,0){2}}
\put ( 28,180) {\line(1,0){4}}
\put ( 30,210) {\line(1,0){2}}
\put ( 30,240) {\line(1,0){2}}
\put ( 30,270) {\line(1,0){2}}
\put ( 30,300) {\line(1,0){2}}
\put ( 28,330) {\line(1,0){4}}
\put ( 30,360) {\line(1,0){2}}
\put ( 30,390) {\line(1,0){2}}
\put ( 30,420) {\line(1,0){2}}
\put ( 15, 27) {0}
\put ( 15,177) {5}
\put ( 10,327) {10}
\put (-10,390) {E(MeV)}
\put( 40, 30.0) {\line(1,0){15}}
\put( 40, 50.5) {\line(1,0){15}}
\put( 40, 82.2) {\line(1,0){15}}
\put( 40,116.1) {\line(1,0){15}}
\put( 40,162.9) {\line(1,0){15}}
\put( 40,223.5) {\line(1,0){15}}
\put( 40,325.8) {\line(1,0){15}}
\put( 35,-10.0) {$\begin{array}{c} \frac{3}{2}^+ \\ \\ 138 \\ \mbox{} \end{array}$}
\put( 57, 27.0) {$\frac{3}{2}^+$}
\put( 57, 47.5) {$\frac{5}{2}^+$}
\put( 57, 79.2) {$\frac{7}{2}^+$}
\put( 57,113.1) {$\frac{9}{2}^+$}
\put( 57,159.9) {$\frac{11}{2}^+$}
\put( 55,220.5) {($\frac{13}{2}^+$)}
\put( 55,322.8) {($\frac{15}{2}^+$)}
\put( 80,113.7) {\line(1,0){15}}
\put( 80,139.8) {\line(1,0){15}}
\put( 80,146.4) {\line(1,0){15}}
\put( 80,189.9) {\line(1,0){15}}
\put( 80,210.9) {\line(1,0){15}}
\put( 80,268.8) {\line(1,0){15}}
\put( 80,312.0) {\line(1,0){15}}
\put( 80,389.4) {\line(1,0){15}}
\put( 75,-10.0) {$\begin{array}{c} \frac{1}{2}^- \\ \\ 148 \\ 0.54 \end{array}$}
\put( 97,110.7) {$\frac{1}{2}^-$}
\put( 97,132.8) {$\frac{3}{2}^-$}
\put( 97,147.4) {$\frac{5}{2}^-$}
\put( 97,186.9) {$\frac{7}{2}^-$}
\put( 97,207.9) {$\frac{9}{2}^-$}
\put( 95,265.8) {($\frac{11}{2}^-$)}
\put( 95,309.0) {($\frac{13}{2}^-$)}
\put( 95,386.4) {($\frac{15}{2}^-$)}
\put(120,113.7) {\line(1,0){15}}
\put(120,142.2) {\line(1,0){15}}
\put(120,170.4) {\line(1,0){15}}
\put(120,218.1) {\line(1,0){15}}
\put(120,269.4) {\line(1,0){15}}
\put(115,-10.0) {$\begin{array}{c} \frac{1}{2}^+ \\ \\ 216 \\ 1.86 \end{array}$}
\put(137,110.7) {$\frac{1}{2}^+$}
\put(137,139.2) {$\frac{5}{2}^+$}
\put(137,167.4) {$\frac{3}{2}^+$}
\put(137,215.1) {$\frac{9}{2}^+$}
\put(135,266.4) {($\frac{7}{2},\frac{11}{2}$)$^+$}
\put(160,165.6) {\line(1,0){15}}
\put(160,192.9) {\line(1,0){15}}
\put(160,226.5) {\line(1,0){15}}
\put(160,277.2) {\line(1,0){15}}
\put(160,321.0) {\line(1,0){15}}
\put(155,-10.0) {$\begin{array}{c} \frac{5}{2}^+ \\ \\ 131 \\ \mbox{} \end{array}$}
\put(177,162.6) {$\frac{5}{2}^+$}
\put(175,189.9) {($\frac{7}{2}^+$)}
\put(177,223.5) {$\frac{9}{2}^{(+)}$}
\put(175,274.2) {($\frac{11}{2}^+$)}
\put(175,318.0) {($\frac{13}{2}^+$)}
\put(200,171.6) {\line(1,0){15}}
\put(200,200.7) {\line(1,0){15}}
\put(200,204.6) {\line(1,0){15}}
\put(200,277.8) {\line(1,0){15}}
\put(195,-10.0) {$\begin{array}{c} \frac{1}{2}^- \\ \\ 174 \\ -2.85 \end{array}$}
\put(217,168.6) {$\frac{3}{2}^-$}
\put(217,192.7) {$\frac{1}{2}^-$}
\put(217,206.6) {$\frac{7}{2}^-$}
\put(215,274.8) {($\frac{5}{2}^-$)}
\put(240,196.5) {\line(1,0){15}}
\put(240,203.1) {\line(1,0){15}}
\put(240,217.8) {\line(1,0){15}}
\put(240,234.0) {\line(1,0){15}}
\put(235,-10.0) {$\begin{array}{c} \frac{3}{2}^+ \\ \\ 60 \\ \mbox{} \end{array}$}
\put(257,186.5) {$\frac{3}{2}^+$}
\put(255,200.1) {($\frac{3}{2},\frac{5}{2}$)}
\put(255,214.8) {($\frac{7}{2}^+$)}
\put(257,231.0) {$\frac{9}{2}^+$}
\put(280,204.9) {\line(1,0){15}}
\put(280,215.1) {\line(1,0){15}}
\put(280,222.3) {\line(1,0){3}}
\put(286,222.3) {\line(1,0){3}}
\put(292,222.3) {\line(1,0){3}}
\put(280,241.2) {\line(1,0){15}}
\put(275,-10.0) {$\begin{array}{c} \frac{3}{2}^+ \\ \\ 58 \\ \mbox{} \end{array}$}
\put(297,196.9) {$\frac{3}{2}^+$}
\put(295,212.1) {($\frac{5}{2}^+$)}
\put(295,238.2) {($\frac{9}{2}^+$)}
\end{picture}
\caption{Cluster interpretation of the rotational bands in $^{21}$Ne. The
bands are labeled by $K^{P}$ and the values of the rotational
parameter, $B$, and decoupling parameter, $a$.}
\label{Ne21}
\end{figure}

The level scheme corresponding to these assignments is shown in Fig.~\ref{Ne21}. 

\subsection{Summary of assignments into bands and comparison with other
assignments and models}

In Table~\ref{assignmentsNe} we show a summary of our assignments into rotational bands and the corresponding values of the intrinsic energies $\varepsilon $, inertial parameters $B$ and decoupling parameters $a$, together with an estimate of
the error in extracting these values from experiment. We also show in this
table, for comparison with the calculated values, the intrinsic energies $%
\varepsilon_{K}$ and $\varepsilon_{K}^{\prime}$ and the character of the
state: p=particle, h=hole,and v=vibration.

\begin{table}
\centering
\caption[Assignments]{Summary of assignments in $^{21}$Ne}
\label{assignmentsNe}
\vspace{10pt}
\begin{tabular}{lccccccc}
\hline
\noalign{\smallskip}
$\#$ & $K^P$ & $E$ & $B$ & $a$ & $\varepsilon$ 
& $\varepsilon_K$ & $\varepsilon'_K$ \\
&& (keV) & (keV) && (keV) & (MeV) & (MeV) \\
\noalign{\smallskip}
\hline
\noalign{\smallskip}
1 (p) & $3/2^{+}$ &    0 & 138(4)  &           & --207(21) &   0    &  --6.97 \\
2 (h) & $1/2^{-}$ & 2789 & 148(12) &   0.54(6) &  2824(4)  & --2.99 &  --9.96 \\
3 (h) & $1/2^{+}$ & 2794 & 216(8)  &   1.86(4) &  3088(29) & --3.31 & --10.28 \\
4 (p) & $5/2^{+}$ & 4526 & 131(4)  &           &  4199(32) &   4.41 &  --2.56 \\
5 (p) & $1/2^{-}$ & 5690 & 174(12) & --2.85(6) &  5107(58) &   5.32 &  --1.65 \\
6 (v) & $3/2^{+}$ & 5549 &  60(2)  &           &  5459(9)  &   5.67 &  --1.30 \\
7 (v) & $3/2^{+}$ & 5826 &  58(2)  &           &  5735(9)  &   6.17 &  --0.80 \\
\noalign{\smallskip}
\hline
\end{tabular}
\end{table}

The assignments of states of bands 1, 3, 4, 6 agree with the assignments of 
\cite{wheldon}. They disagree for bands 2, 5 and 7. Band 2 is assigned in 
\cite{wheldon} as $K^{P}=3/2^{-}$. This in contrast with experiment for two
reasons: (i) the $5/2^{-}$ member of the supposed $K=1/2^{-}$ band is
missing and (ii) most importantly the $B(E2;3/2^{-}\rightarrow 1/2^{-})$ is
observed to be very large $15(13)$ W.u. $=51.6(447)$ e$^{2}$fm$^{4}$, as
discussed in the following section. The assignment of band 7 to $%
K^{P}=1/2^{+}$ is also in contrast with experiment. The authors of \cite%
{wheldon} have changed the parity of the state at $5690$ keV from $%
J^{P}=1/2^{-}$ \cite{firestone} to $J^{P}=1/2^{+}$.

In summary, there appear to be in $^{21}$Ne three coexisting classes of
rotational bands, classified in the cluster model as single-particle states 
(bands 1, 4 and 5), as single-hole states (bands 2 and 3), and as vibrational states (bands 6 and 7). The CSM provides an excellent description of bands 
1, 2 and 3, and a good description of bands 4, 5, 6 and 7. Its only drawback 
is the non-observation of the band $K^{P}=1/2^{+}$ at 
$\varepsilon_{K}^{\prime}=-5.44$ MeV.
Although one may try to correct this problem by the introduction of \textit{%
ad hoc} terms in the potential, we do not pursue this avenue further in this
paper. We also note that the CSM predicts the neutron separation energy $%
S_{n}(\rm th)=7.01$ MeV, in very good agreement with the experimental value $%
S_{n}(\exp )=6761.14(4)$ keV.

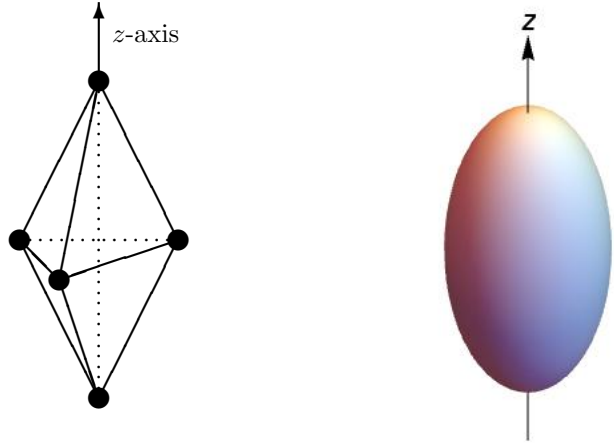
\begin{figure}
\centering
\vspace{15pt}
\begin{minipage}{0.5\linewidth}
\setlength{\unitlength}{1pt}
\begin{picture}(120,200)(0,0)
\thicklines
\put( 45, 85) {\circle*{8}} 
\put( 30,100) {\circle*{8}}
\put( 90,100) {\circle*{8}}
\put( 60,160) {\circle*{8}}
\put( 60, 40) {\circle*{8}}
\put( 65,175) {$z$-axis}
\put( 60,160) {\vector( 0, 1){30}}
\multiput( 30,100)(4,0){15}{\circle*{1}}
\multiput( 60, 40)(0,4){30}{\circle*{1}}
\put( 30,100) {\line( 1, 2){30}}
\put( 30,100) {\line( 1,-1){15}}
\put( 30,100) {\line( 1,-2){30}}
\put( 90,100) {\line(-1, 2){30}}
\put( 90,100) {\line(-1,-2){30}}
\put( 90,100) {\line(-3,-1){45}}
\put( 45, 85) {\line( 1, 5){15}}
\put( 45, 85) {\line( 1,-3){15}}
\end{picture}
\end{minipage}
\begin{minipage}{0.3\linewidth}
\includegraphics[scale=0.5]{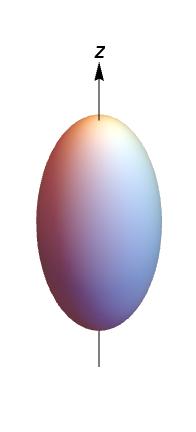}
\end{minipage}
\caption[Ellipsoid]{The bi-pyramidal structure with ${\cal D}_{3h}$ symetry 
and its osculating ellipsoid with ${\cal D}_{\infty h}$ symmetry.}
\label{ellips}
\end{figure}

It is of interest to compare the results of the CSM with those of the
Nilsson model \cite{nilsson}. To this end, we show in the right column of Fig.~\ref{defspNe} the Nilsson intrinsic energies $\varepsilon_{K}$ calculated 
in \cite{howard1}. The Nilsson model produces intrinsic energies $\varepsilon_{K}$
comparable to those of the CSM and in good agreement with data. The enlarged
scale of the Nilsson model relative to the CSM is due to the fact that in
the Nilsson model an harmonic oscillator potential is used while in the CSM
a gaussian potential is used. As in the case of CSM, the drawback of the
Nilsson model is the non-observation of the band $K^{P}=1/2^{+}$ at $%
\varepsilon_{K}=2.82$\ MeV. Also, in the Nilsson model, one cannot
calculate the separation energy, $S_{n}$. The agreement between the
intrinsic energies of the two models arises from the fact that the
bi-pyramid can be inscribed into an ellispsoid with large deformation, as
shown in Fig.~\ref{ellips}.

Our assignments suggest the following spherical shell model interpretration:
the positive parity bands 1, 3 and 4 arise mostly from $(2s1d)^5$ configurations, 
the negative parity band 2 arises from $(2s1d)^6 (1p)^{-1}$ configuration and the negative parity band 5 from $(2s1d)^4(1f)$ configuration. Since the $A'_1$ 
vibrations in $^{20}$Ne arise in the shell model from $4p-4h$ configurations, the two vibrational bands 6 and 7 in $^{21}$Ne arise from configurations 
$(2s1d)^5 \times (4p-4h)$. Both the $A'_1$ bands in $^{20}$Ne and 
$A'_1 \otimes K^P = 3/2^+$ have energies of about 6 MeV which is the energy
of the expected $4p-4h$ states.

\subsection{Cluster interpretation of the observed bands}

The particle and vibrational bands of Table~\ref{assignmentsNe} can be identified as states of the cluster structure $^{20}$Ne+n. They arise from coupling of the
rotational bands of $^{20}$Ne denoted by $\Gamma :K_{c}^{P_{c}}(E_{c})$ with
single-particle states denoted by $\Omega :K_{p}^{P_{p}}(E_{p})$. The values
of $K$, the total projection on the symmetry axis, and $P$, the total parity
are given by%
\begin{equation}
K=\left\vert K_{c}\pm K_{p}\right\vert ~, \qquad \qquad P=P_{c}P_{p} ~.
\end{equation}%
(If $K_{c}\neq 0$, there are two values of $K$). The three particle bands $%
K^{P}=3/2^{+}(0.0),$ $K^{P}=5/2^{+}(4526),$ $K^{P}=1/2^{-}(5690)$ are the
corresponding particle intrinsic states coupled to the ground state of $^{20}
$Ne, $A_{1}^{\prime }:K^{P}=0^{+}(0.0)$. The two vibrational bands $%
K^{P}=3/2^{+}(5549)$ and $K^{P}=3/2^{+}(5826)$ are the particle state $%
K^{P}=3/2^{+}$ coupled to the vibrations of $^{20}$Ne, $A_{1}^{\prime
}:K^{P}=0^{+}(6.72)$ and $A_{1}^{\prime}=0^{+}(7.19)$. On the basis of
Table~\ref{gsvib}, one would expect seven more vibrational bands. Although there is
some evidence for additional bands with intrinsic energies higher than $6$
MeV, the analysis of bands at these energies is made very difficult by the
high density of states and we will not pursue it further in this paper. The
two hole bands of Table~\ref{assignmentsNe}, $K^{P}=1/2^{-}(2789)$, $K^{P}=1/2^{+}(2796)$,
forming a parity doublet, can be identified as states of the cluster
structure $^{19}$Ne+2n. This identification is strongly supported by the
experimental spectrum of $^{19}$Ne, the ground state of which is a parity
doublet $K^{P}=1/2^{+}(0.0),$ $K^{P}=1/2^{-}(275)$. In summary, it appears
that in $^{21}$Ne there is the coexistence of two cluster structures, $^{20}$%
Ne+n and $^{19}$Ne+2n.

\section{Electromagnetic transition rates}
\label{sec5}

The reduced probabilities of electromagnetic transitions in $k\alpha +x$ nuclei can be calculated using Eq.~(25) of \cite{dellarocca2} as
\begin{eqnarray}
&& B(\lambda ;\Omega',K',J'^{P'} \rightarrow \Omega,K,J^P)
\nonumber\\
&& \qquad =\Big| \left\langle J',K',\lambda,K-K'|J,K \right\rangle
\left( \delta_{v,v'} G_{\lambda}(\Omega,\Omega')
+ \delta_{\Omega,\Omega'} G_{\lambda,c} \right) 
\nonumber\\ 
&& \qquad \qquad + (-1)^{J+K} 
\left\langle J',K',\lambda,-K-K'|J,-K \right\rangle 
\nonumber\\
&& \qquad \qquad \times 
\left( \delta _{v,v'} \tilde{G}_{\lambda}(\Omega,-\Omega')
+ \delta_{\Omega,-\Omega'} G_{\lambda,c} \right) \Big|^2
\label{blam}
\end{eqnarray}
We note that the second term in Eq.~(\ref{blam}) contributes only in the case $%
\lambda \geq K+K'$. Here $G_{\lambda}(\Omega,\Omega')$
represents the contribution of the single particle and $G_{\lambda,c}$ the
contribution of the cluster. Similarly, electric and magnetic multipole
moments can be calculated using Eq.~(28) of \cite{dellarocca2}%
\begin{eqnarray}
Q^{(\lambda )}(K,J^P) &=& \sqrt{\frac{16\pi}{2\lambda+1}} 
\left\langle J,K,\lambda,0|J,K \right\rangle 
\left\langle J,J,\lambda,0|J,J \right\rangle
\left( G_{\lambda}(\Omega,\Omega) + G_{\lambda,c} \right) ~,
\nonumber\\
\mu^{(\lambda)}(K,J^P) &=& \sqrt{\frac{4\pi}{2\lambda+1}} 
\left\langle J,K,\lambda,0|J,K \right\rangle 
\left\langle J,J,\lambda,0|J,J \right\rangle 
\left(G_{\lambda}(\Omega,\Omega) + G_{\lambda,c} \right) ~.
\nonumber\\
\mbox{}
\label{qmu}
\end{eqnarray}
Here again $G_{\lambda}(\Omega,\Omega)$ represents the contribution of
the single particle and $G_{\lambda,c}$ of the cluster. In 
$^{21}$Ne, the single particle is a neutron and thus it does not contribute
to electric transitions, except for $E1$ transitions affected by center-of-mass corrections as given in Eq.~(32) of \cite{dellarocca2}. The cluster contribution to electric transitions $E2$, $E3$, $E4$ is given by the ${\cal D}_{3h}$
symmetry as in Sect.~2.2.4 of \cite{bijker5}. Note that the cluster
calculation of electric transtions does not require the use of effective
charges. Magnetic transitions are dominated by the single-particle
contribution. For in-band transitions and magnetic moments, there is also a
cluster contribution. 

\subsection{$K^{P}=\tfrac{3}{2}^{+}(0) \rightarrow K^{P}=\tfrac{3}{2}^{+}(0)$}

Several transition rates have been measured for this band. The $B(E2)$
values and quadrupole moments $Q^{(2)}$ can be simply calculated as%
\begin{eqnarray}
B(E2;\tfrac{3}{2},J'^+ \rightarrow \tfrac{3}{2},J^+) 
&=& Q_{0}^{2} \frac{5}{16\pi} 
\left\langle J^{\prime},\tfrac{3}{2},2,0|J,\tfrac{3}{2} \right\rangle^{2} ~,
\nonumber\\
Q^{(2)}(\tfrac{3}{2},J^+) &=& Q_{0} 
\left\langle J,\tfrac{3}{2},2,0|J,\tfrac{3}{2} \right\rangle 
\left\langle J,J,2,0|J,J \right\rangle ~.
\label{be33}
\end{eqnarray}
The value of the intrinsic quadrupole moment in $^{21}$Ne,  $Q_{0}=G_{2,c}\sqrt{16\pi/5}$, can be determined from the quadrupole transition $B(E2;\tfrac{3}{2},\tfrac{5}{2}^+ \rightarrow \tfrac{3}{2},\tfrac{3}{2}^+)$ as $Q_{0}(^{21}\mbox{Ne})=49.4(20)$ efm$^{2}$, in remarkable agreement with the value obtained for $^{20}$Ne \cite{bijker5}, $Q_{0}(^{20}\mbox{Ne})=52.5(21)$ efm$^{2}$. With the value of $Q_{0}=49.4$ efm$^{2}$ we calculate all $B(E2)$ values and quadrupole moments as given in Tables~\ref{BEM1} and \ref{mom1}. The
asterisk in this and the following tables indicates that the value is used to
extract the intrinsic matrix elements.

Similarly, the $B(M1)$ values and magnetic moments are given by%
\begin{eqnarray}
B(M1;\tfrac{3}{2},J'^+ \rightarrow \tfrac{3}{2},J^+) 
&=& \left\langle J',\tfrac{3}{2},1,0|J,\tfrac{3}{2} \right\rangle^{2} 
\left\vert G_{1}(\tfrac{3}{2}^{+}) \right\vert^{2} ~,
\nonumber\\
\mu^{(1)}(\tfrac{3}{2},J^+) &=& \sqrt{\frac{4\pi}{3}} 
\left\langle J,\tfrac{3}{2},1,0|J,\tfrac{3}{2} \right\rangle 
\left\langle J,J,1,0|J,J\right\rangle 
\nonumber\\
&& \qquad \qquad \left[ G_{1}(\tfrac{3}{2}^{+})+G_{1R} \right] ~,
\label{bm33}
\end{eqnarray}
where $G_{1R}$ represents an additional contribution of the cluster. From
the $B(M1;\tfrac{3}{2},\tfrac{7}{2}^+ \rightarrow \tfrac{3}{2},\tfrac{5}{2}^+)$ 
we extract the value $| G_{1}(\tfrac{3}{2}^{+}) | =0.85(4)$ $\mu_{N}$. 
From this value we can calculate all $B(M1)$ values given in Table~\ref{BEM1}. 
For the magnetic moment using $G_{1}(\tfrac{3}{2}^{+})=-0.85$ $\mu _{N}$ and adding $G_{1R}=Z/A=0.50$ $\mu_{N}$ we obtain the value of Table~\ref{mom1}. The Tables~\ref{BEM1} and \ref{mom1} show that the band $K^{P}=\tfrac{3}{2}^{+}(0)$ is an almost perfect rotational band.

\begin{table}
\centering
\caption[BEM1]{In-band $B(E2)$ values in e$^{2}$fm$^{4}$ and 
$B(M1)$ values in $\mu_{N}^{2}$ for the $K^{P}=\tfrac{3}{2}^{+}(0)$ band. 
The theoretical values were obtained with $Q_{0}=49.4$ efm$^{2}$ and $G_{1}(\tfrac{3}{2}^{+})=-0.85$ $\mu_N$.}
\label{BEM1}
\vspace{10pt}
\begin{tabular}{crcrllll}
\hline
\noalign{\smallskip}
&& && \multicolumn{2}{c}{$B(E2)$} & \multicolumn{2}{c}{$B(M1)$} \\ 
$E_{\gamma}^{\exp}$ (keV) & $J'^{P'} \;$ & $\rightarrow$ & $J^P \;$ 
& Exp & Th & Exp & Th \\ 
\noalign{\smallskip}
\hline
\noalign{\smallskip}
 351 & $ 5/2^{+}$ & $\rightarrow$ & $ 3/2^{+}$ 
& 83.6(61) & 83.2$^{\ast}$ & 0.1275(25) & 0.19 \\ 
1394 & $ 7/2^{+}$ & $\rightarrow$ & $ 5/2^{+}$ 
& 37.8(137) & 52.0 & 0.2615(21) & 
0.26$^{\ast}$ \\ 
1122 & $ 9/2^{+}$ & $\rightarrow$ & $ 7/2^{+}$ & 31.0(172) & 34.0 & 0.43(5) & 0.29 \\ 
1566 & $11/2^{+}$ & $\rightarrow$ & $ 9/2^{+}$ & 20.6(130) & 23.8 & 0.36(7) & 0.31 \\ 
2015 & $13/2^{+}$ & $\rightarrow$ & $11/2^{+}$ & & 17.4 & & 0.32 \\ 
3409 & $15/2^{+}$ & $\rightarrow$ & $13/2^{+}$ & & 13.3 & & 0.33 \\ 
\noalign{\smallskip}
1745 & $ 7/2^{+}$ & $\rightarrow$ & $ 3/2^{+}$ & 32.0(27) & 34.7 & & \\ 
2516 & $ 9/2^{+}$ & $\rightarrow$ & $ 5/2^{+}$ & 54.7(76) & 52.0 & & \\ 
2688 & $11/2^{+}$ & $\rightarrow$ & $ 7/2^{+}$ &          & 61.8 & & \\ 
3581 & $13/2^{+}$ & $\rightarrow$ & $ 9/2^{+}$ &          & 67.9 & & \\ 
5424 & $15/2^{+}$ & $\rightarrow$ & $11/2^{+}$ &          & 72.0 & & \\
\noalign{\smallskip}
\hline
\end{tabular}
\end{table}

\begin{table}
\centering
\caption[Moments]{Spectroscopic quadrupole moment and magnetic moment of the $J^P=\tfrac{3}{2}^{+}$ ground state.}
\label{mom1}
\vspace{10pt}
\begin{tabular}{cccc}
\hline
\noalign{\smallskip}
& Exp & Th & \\ 
\noalign{\smallskip}
\hline
\noalign{\smallskip}
  $Q^{(2)}(\tfrac{3}{2},\tfrac{3}{2}^+)$ & $+10.3(8)$ & $+9.9$ & efm$^{2}$ \\
\noalign{\smallskip}
$\mu^{(1)}(\tfrac{3}{2},\tfrac{3}{2}^+)$ & $-0.661797(5)$ & $-0.43$ & $\mu_N$ \\
\noalign{\smallskip}
\hline
\end{tabular}
\end{table}

\subsection{$K^{P}=\tfrac{1}{2}^{-}(2789) \rightarrow K^{P}=\tfrac{1}{2}^{-}(2789)$}

$B(E2)$ values and quadrupole moments for this band can be calculated with 
\begin{eqnarray}
B(E2;\tfrac{1}{2},J'^- \rightarrow \tfrac{1}{2},J^-) &=& Q_{0}^{2} \frac{5}{16\pi} \left\langle J',\tfrac{1}{2},2,0|J,\tfrac{1}{2} \right\rangle^{2} ~,
\nonumber\\
Q^{(2)}(\tfrac{1}{2},J^-) &=& Q_{0} 
\left\langle J,\tfrac{1}{2},2,0|J,\tfrac{1}{2} \right\rangle 
\left\langle J,J,2,0|J,J \right\rangle ~.
\label{be11}
\end{eqnarray}
Although in general for $K=\tfrac{1}{2}$ bands both terms in Eq.~(\ref{blam}) contribute, in this special case in which there is no single-particle contribution, 
only the first term remains as in Eq.~(\ref{be11}). The results of the
calculation for $B(E2)$ values are shown in Table~\ref{BEM2}, where they are compared
with the only available experimental value. The large value of 
$B(E2;\tfrac{1}{2},\tfrac{3}{2}^{-} \rightarrow \tfrac{1}{2},\tfrac{1}{2}^{-})$ is strong evidence for the assignment of this band to $K^P=\tfrac{1}{2}^{-}$, in contrast with the assignment of \cite{wheldon}. 
Similarly one can calculate $B(M1)$ values and magnetic moments for this band as 
\begin{eqnarray}
B(M1;\tfrac{1}{2},J'^- \rightarrow \tfrac{1}{2},J^-) &=& 
\left\langle J',\tfrac{1}{2},1,0|J,\tfrac{1}{2} \right\rangle^{2} 
\left\vert G_{1}(\tfrac{1}{2}^{-}) \right\vert^{2} ~,
\nonumber\\
\mu^{(1)}(\tfrac{1}{2},J^-) &=& \sqrt{\frac{4\pi}{3}} 
\left\langle J,\tfrac{1}{2},1,0|J,\tfrac{1}{2} \right\rangle 
\left\langle J,J,1,0|J,J \right\rangle 
\nonumber\\
&& \qquad \qquad \left[ G_{1}(\tfrac{1}{2}^{-})+G_{1R} \right] ~.
\label{bm11}
\end{eqnarray}

\begin{table}
\centering
\caption[BEM2]{In-band $B(E2)$ values in e$^{2}$fm$^{4}$ and 
$B(M1)$ values in $\mu_{N}^2$) for the $K^{P}=\tfrac{1}{2}^{-}(2789)$ band. 
The theoretical values were obtained with $Q_{0}=49.4$ efm$^{2}$ 
and $| G_{1}(\tfrac{1}{2}^{-}) |= 0.993$ $\mu_N$.}
\label{BEM2}
\vspace{10pt}
\begin{tabular}{crcrlrll}
\hline
\noalign{\smallskip}
&& && \multicolumn{2}{c}{$B(E2)$} & \multicolumn{2}{c}{$B(M1)$} \\ 
$E_{\gamma}^{\exp}$ (keV) & $J'^{P'} \;$ & $\rightarrow$ & $J^P \;$ 
& Exp & Th & Exp & Th \\ 
\noalign{\smallskip}
\hline
\noalign{\smallskip}
 874 & $ 3/2^{-}$ & $\rightarrow$ & $ 1/2^{-}$ 
 & 51.6(447) & 48.6 & 0.329(43) & 0.33$^{\ast}$ \\ 
 221 & $ 5/2^{-}$ & $\rightarrow$ & $ 3/2^{-}$ && 13.9 && 0.40 \\ 
1450 & $ 7/2^{-}$ & $\rightarrow$ & $ 5/2^{-}$ &&  6.9 && 0.43 \\ 
 699 & $ 9/2^{-}$ & $\rightarrow$ & $ 7/2^{-}$ &&  4.2 && 0.44 \\ 
1928 & $11/2^{-}$ & $\rightarrow$ & $ 9/2^{-}$ &&  2.8 && 0.45 \\ 
1440 & $13/2^{-}$ & $\rightarrow$ & $11/2^{-}$ &&  2.0 && 0.46 \\ 
2582 & $15/2^{-}$ & $\rightarrow$ & $13/2^{-}$ &&  1.5 && 0.46 \\ 
\noalign{\smallskip} 
1095 & $ 5/2^{-}$ & $\rightarrow$ & $ 1/2^{-}$ && 48.6 && \\ 
1671 & $ 7/2^{-}$ & $\rightarrow$ & $ 3/2^{-}$ && 62.4 && \\ 
2149 & $ 9/2^{-}$ & $\rightarrow$ & $ 5/2^{-}$ && 69.4 && \\ 
2627 & $11/2^{-}$ & $\rightarrow$ & $ 7/2^{-}$ && 73.6 && \\ 
3368 & $13/2^{-}$ & $\rightarrow$ & $ 9/2^{-}$ && 76.4 && \\ 
4002 & $15/2^{-}$ & $\rightarrow$ & $11/2^{-}$ && 78.4 && \\
\noalign{\smallskip}
\hline
\end{tabular}
\end{table}

We remark that the measured value of the $B(E2;\tfrac{1}{2},\tfrac{3}{2}^{-} \rightarrow \tfrac{1}{2},\tfrac{1}{2}^{-})$ is strong evidence for the cluster structure of $^{21}$Ne as $^{20}$Ne+n because it agrees within experimental error with the calculated value with $Q_{0}=49.4$ efm$^{2}$, which in turn agrees with $Q_{0}$ in $^{20}$Ne. No new parameter is involved in the calculation of the quadrupole transitions. 
The value of $| G_{1}(\tfrac{1}{2}^{-}) |$ is determined from the measured value 
of $B(M1;\tfrac{1}{2},\tfrac{3}{2}^{-} \rightarrow \tfrac{1}{2},\tfrac{1}{2}^{-})$ 
to be $0.993$ $\mu _{N}$. The $B(M1)$ values are given in Table~\ref{BEM2}.

\subsection{$K^{P}=\tfrac{1}{2}^{-}(2789) \rightarrow K^{P}=\tfrac{3}{2}^{+}(0)$}

Several out of band $E1$ transitions rates from $K^{P}=\tfrac{1}{2}^{-}(2789)$ 
to the ground state band $K^{P}=\tfrac{3}{2}^{+}(0)$ have been measured \cite{firestone}. $E1$ transition rates are difficult to calculate because 
they vanish in $N=Z$ nuclei due to isospin considerations. In $^{21}$Ne, 
they arise from center-of-mass corrections \cite{dellarocca2}. Within the cluster model, they can be parametrized in terms of an effective dipole moment $Q_{1m}$ $(m=0,\pm 1)$. For the $\Delta K=1$ transitions considered here, they can be calculated using
\begin{equation}
B(E1;\tfrac{1}{2},J'^- \rightarrow \tfrac{3}{2},J^+) = 2 \left( Q_{11} \right)^{2} \frac{3}{4\pi} \left\langle J',\tfrac{1}{2},1,1|J,\tfrac{3}{2} \right\rangle^{2},
\end{equation}
the factor of 2 arising from the sum over $Q_{1,+1}$ and $Q_{1,-1}$. By
fitting the value of the dipole moment $Q_{11}$ to the dipole transition 
$B(E1;\tfrac{1}{2},\tfrac{5}{2}^{-} \rightarrow \tfrac{3}{2},\tfrac{5}{2}^{+})$ 
we obtain $Q_{11}=3.45 \times 10^{-2}$ efm. Table~\ref{BEM3} shows that 
the calculated values are in excellent agreement with experiment.

\begin{table}
\centering
\caption[BEM3]{$K^{P}=\tfrac{1}{2}^{-}$(2789) to $K^{P}=\tfrac{3}{2}^{+}$(0) 
interband $10^{4} \times B(E1)$ values in e$^{2}$fm$^{2}$ 
and $B(E3)$ values in e$^{2}$fm$^{6}$. The theoretical values were obtained 
with $Q_{11}=3.45 \times 10^{-2}$ efm and $Q_{32}=16.3$ efm$^3$.}
\label{BEM3}
\vspace{10pt}
\begin{tabular}{crcrll}
\hline
\noalign{\smallskip}
&& && \multicolumn{2}{c}{$B(E1)$} \\ 
$E_{\gamma}^{\exp}$ (keV) & $J'^{P'} \;$ & $\rightarrow$ & $J^P \;$ & Exp & Th \\ 
\noalign{\smallskip}
\hline
\noalign{\smallskip}
 2789  & $ 1/2^{-}$ & $\rightarrow$ & $ 3/2^{+}$ &          & 5.69 \\ 
 3312  & $ 3/2^{-}$ & $\rightarrow$ & $ 5/2^{+}$ & 1.09(12) & 3.41 \\ 
 3663  & $ 3/2^{-}$ & $\rightarrow$ & $ 3/2^{+}$ &          & 2.28 \\ 
 2139  & $ 5/2^{-}$ & $\rightarrow$ & $ 7/2^{+}$ &          & 2.71 \\ 
 3533  & $ 5/2^{-}$ & $\rightarrow$ & $ 5/2^{+}$ & 2.6(3)   & 2.60$^{\ast}$ \\ 
 3884  & $ 5/2^{-}$ & $\rightarrow$ & $ 3/2^{+}$ & 0.65(11) & 0.38 \\ 
 2467  & $ 7/2^{-}$ & $\rightarrow$ & $ 9/2^{+}$ &          & 2.37 \\ 
 3589  & $ 7/2^{-}$ & $\rightarrow$ & $ 7/2^{+}$ &          & 2.71 \\ 
 4983  & $ 7/2^{-}$ & $\rightarrow$ & $ 5/2^{+}$ &          & 0.61 \\ 
 1600  & $ 9/2^{-}$ & $\rightarrow$ & $11/2^{+}$ &          & 2.17 \\ 
 3166  & $ 9/2^{-}$ & $\rightarrow$ & $ 9/2^{+}$ & 2.35(29) & 2.76 \\ 
 4288  & $ 9/2^{-}$ & $\rightarrow$ & $ 7/2^{+}$ & 1.13(29) & 0.76 \\ 
 1513  & $11/2^{-}$ & $\rightarrow$ & $13/2^{+}$ &          & 2.04 \\ 
 3528  & $11/2^{-}$ & $\rightarrow$ & $11/2^{+}$ &          & 2.78 \\ 
 5094  & $11/2^{-}$ & $\rightarrow$ & $ 9/2^{+}$ &          & 0.86 \\ 
(-456) & $13/2^{-}$ & $\rightarrow$ & $15/2^{+}$ &          & 1.95 \\ 
 2953  & $13/2^{-}$ & $\rightarrow$ & $13/2^{+}$ &          & 2.80 \\ 
 4968  & $13/2^{-}$ & $\rightarrow$ & $11/2^{+}$ &          & 0.94 \\
\noalign{\smallskip}
\hline
\noalign{\smallskip}
&& && \multicolumn{2}{c}{$B(E3)$} \\ 
$E_{\gamma}^{\exp}$ (keV) & $J'^{P'} \;$ & $\rightarrow$ & $J^P \;$ & Exp & Th \\ 
\noalign{\smallskip}
\hline
\noalign{\smallskip}
2438 & $1/2^{-}$ & $\rightarrow$ & $5/2^{+}$ & 340(183) & 380 \\
\noalign{\smallskip}
\hline
\end{tabular}
\end{table}

The extraction of the electric dipole moment in cluster models has been the
subject of a recent investigation \cite{wheldon}. The dipole moment $D$ of 
\cite{wheldon} is related to $Q_{11}$ by 
\begin{equation}
2(Q_{11})^{2}=(D)^{2} ~.
\end{equation}%
We obtain from $Q_{11}=3.45\times 10^{-2}$ efm, $D=0.0488(26)$ efm. Wheldon 
\textit{et al.} extract the dipole moment assuming that the $J^{P}=\tfrac{9}{2}^{-}$ state is part of a $K^{P}=\tfrac{3}{2}^{-}$ band 
obtaining $D=0.0337(79)$ efm (from the transition $\tfrac{9}{2}^{-}\rightarrow \tfrac{7}{2}^{+}$) and $D=0.097(19)$ efm (from $\tfrac{9}{2}^{-}\rightarrow
\tfrac{9}{2}^{+}$) \cite{wheldon}. These values agree only at the $3\sigma $ level. 
If we extract the dipole moment assuming the $J^{P}=\tfrac{9}{2}^{-}$ state to be part of a $K^{P}=\tfrac{1}{2}^{-}$ band we obtain $D=0.0450(27)$ (from $\tfrac{9}{2}^{-} \rightarrow \tfrac{9}{2}^{+}$) and $D=0.0596(77)$ efm (from $\tfrac{9}{2}^{-} \rightarrow \tfrac{7}{2}^{+}$). These values are
within experimental error thus confirming our assignment. 

${\cal D}_{3h}$ symmetry implies large $E2$, $E3$ and $E4$ transition rates 
\cite{bijker1}. In \cite{bijker5} $E3$ transitions in $^{20}$Ne have been
analyzed. The two observed $\Delta K=2$\ transitions have been analyzed by
using
\begin{equation}
B(E3;\tfrac{1}{2},J'^- \rightarrow \tfrac{3}{2},J^+) = 2 \left( Q_{32} \right)^{2} 
\left\langle J',-\tfrac{1}{2},3,2|J,\tfrac{3}{2} \right\rangle^{2} ~,
\end{equation}%
where $Q_{32}$ is the octupole moment. The $B(E3)$ value in Table~\ref{BEM3} 
was obtained by using the same value for the octupole moment as in the 
analysis of $E3$ transitions in $^{20}$Ne, $Q_{32}=16.3$ efm$^{3}$ \cite{bijker5}. 
The large experimental value of the octupole transition $B(E3;\tfrac{1}{2},\tfrac{1}{2}^{-} \rightarrow \tfrac{3}{2},\tfrac{5}{2}^{+}) =13(7)$ W.u. $=340(183)$ e$^{2}$fm$^{6}$ is further evidence of the cluster structure of $^{21}$Ne. 

\subsection{$K^{P}=\tfrac{1}{2}^{-}(5690) \rightarrow K^{P}=\tfrac{1}{2}^{-}(5690)$}

No transition rates have been measured for this band. However, the observed
transition between the $J^{P}=\tfrac{1}{2}^{-}$ and $\tfrac{3}{2}^{-}$ members of
this band strongly support our interpretation of this band as $K^{P}=\tfrac{1}{2}^{-}$. The calculated values with $Q_{0}=49.4$ efm$^{2}$ and with the estimated value $\left\vert G_{1}(\tfrac{1}{2}^{-})\right\vert=1.0$ 
$\mu _{N}$ are given in Table~\ref{BEM5}.

\begin{table}
\centering
\caption[BEM5]{In-band $B(E2)$ values in e$^{2}$fm$^{4}$ and 
$B(M1)$ values in $\mu_{N}^{2}$ for the $K^{P}=\tfrac{3}{2}^{-}(4725)$ band. 
The theoretical values were obtained with $Q_{0}=49.4$ efm$^{2}$ and 
$|G_{1}(\tfrac{1}{2}^{-})|= 1.0$ $\mu_N$. $E_{\gamma}^{\exp}$ values are not 
given here because of the uncertainty in the assignment of states.}
\label{BEM5}
\vspace{10pt}
\begin{tabular}{rcrllll}
\hline
\noalign{\smallskip}
&&& \multicolumn{2}{c}{$B(E2)$} & \multicolumn{2}{c}{$B(M1)$} \\ 
$J'^{P'} \;$ & $\rightarrow$ & $J^P \;$ 
& Exp & Th & Exp & Th \\ 
\noalign{\smallskip}
\hline
\noalign{\smallskip}
$3/2^{-}$ & $\rightarrow$ & $1/2^{-}$ && 48.6 && 0.33 \\
$5/2^{-}$ & $\rightarrow$ & $3/2^{-}$ && 13.9 && 0.40 \\
$7/2^{-}$ & $\rightarrow$ & $5/2^{-}$ &&  6.9 && 0.43 \\ 
$9/2^{-}$ & $\rightarrow$ & $7/2^{-}$ &&  4.2 && 0.44 \\ 
\noalign{\smallskip}
$5/2^{-}$ & $\rightarrow$ & $1/2^{-}$ && 48.6 && \\
$7/2^{-}$ & $\rightarrow$ & $3/2_{-}$ && 62.4 && \\ 
$9/2^{-}$ & $\rightarrow$ & $5/2^{-}$ && 69.4 && \\
\noalign{\smallskip}
\hline
\end{tabular}
\end{table}

\subsection{$K^{P}=\tfrac{1}{2}^{+}(2794) \rightarrow K^{P}=\tfrac{1}{2}^{+}(2794)$ and 
$K^{P}=\tfrac{1}{2}^{+}(2794) \rightarrow K^{P}=\tfrac{3}{2}^{+}(0)$}

In band $B(E2)$ values and quadrupole moments are calculated using Eq.~(\ref{be11}). The same comments apply here as in the paragraph after these
equations. $B(M1)$ values and magnetic moments are calculated with Eq.~(\ref{bm11}). The results are given in Table~\ref{BEM6}. Only one $B(M1)$ value 
and no $B(E2)$ values are available. Therefore no direct test of the cluster calculation can be made.

\begin{table}
\centering
\caption[BEM6]{In-band $B(E2)$ values in e$^{2}$fm$^{4}$ and 
$B(M1)$ values in $\mu_{N}^{2}$ for the $K^{P}=\tfrac{1}{2}^{+}$(2794) band. 
The theoretical values were obtained with $Q_{0}=49.4$ efm$^{2}$ and 
$| G_{1}(\tfrac{1}{2}^{+}) |=0.193$ $\mu_N$.}
\label{BEM6}
\vspace{10pt}
\begin{tabular}{crcrllll}
\hline
\noalign{\smallskip}
&& && \multicolumn{2}{c}{$B(E2)$} & \multicolumn{2}{c}{$B(M1)$} \\ 
$E_{\gamma}^{\exp}$ (keV) & $J'^{P'} \;$ & $\rightarrow$ & $J^P \;$ 
& Exp & Th & Exp & Th \\ 
\noalign{\smallskip}
\hline
\noalign{\smallskip}
  1890  & $ 3/2^{+}$ & $\rightarrow$ & $1/2^{+}$ && 48.6 
& 0.0124(39) & 0.013$^{\ast}$ \\ 
 (-948) & $ 5/2^{+}$ & $\rightarrow$ & $3/2^{+}$ && 13.9 && 0.015 \\ 
  4246  & $ 7/2^{+}$ & $\rightarrow$ & $5/2^{+}$ &&  6.9 && 0.016 \\ 
(-1715) & $ 9/2^{+}$ & $\rightarrow$ & $7/2^{+}$ &&  4.2 && 0.017 \\ 
        & $11/2^{+}$ & $\rightarrow$ & $9/2^{+}$ &&  2.8 && 0.018 \\ 
\noalign{\smallskip}
 942 & $ 5/2^{+}$ & $\rightarrow$ & $1/2^{+}$ && 48.6 && \\ 
3298 & $ 7/2^{+}$ & $\rightarrow$ & $3/2^{+}$ && 62.4 && \\ 
2531 & $ 9/2^{+}$ & $\rightarrow$ & $5/2^{+}$ && 69.4 && \\ 
     & $11/2^{+}$ & $\rightarrow$ & $7/2^{+}$ && 73.6 && \\
\noalign{\smallskip}
\hline
\end{tabular}
\end{table}

$M1$ transitions from the band $K^{P}=\tfrac{1}{2}^{+}(2794)$ to the ground state
band $K^{P}=\tfrac{3}{2}^{+}(0)$ can be calculated by 
\begin{equation}
B(M1;\tfrac{1}{2},J'^+ \rightarrow \tfrac{3}{2},J^+) = 
\left\langle J',\tfrac{1}{2},1,1|J,\tfrac{3}{2} \right\rangle^{2} 
\left\vert G_{1}(\tfrac{1}{2}^{+},\tfrac{3}{2}^{+}) \right\vert^2 ~.
\end{equation}%
The value of $G_{1}(\tfrac{1}{2}^{+},\tfrac{3}{2}^{+})$ can be fitted to the transition 
$B(M1;\tfrac{1}{2},\tfrac{3}{2}^+ \rightarrow \tfrac{3}{2},\tfrac{3}{2}^+)$ 
to give $| G_{1}(\tfrac{1}{2}^{+},\tfrac{3}{2}^{+}) | =0.180$ $\mu_{N}$. 
$E2$ transitions to the ground state vanish in the cluster model since it 
is a neutron transition between two different intrinsic states. 
The results are given in Table~\ref{BEM7}.

\begin{table}
\centering
\caption[BEM7]{$K^{P}=\tfrac{1}{2}^{+}$(2794) to $K^{P}=\tfrac{3}{2}^{+}$(0) 
interband $B(E2)$ values in e$^{2}$fm$^{4}$ and B(M1) values in $\mu_{N}^{2}$. 
The theoretical values were obtained with $Q_{0}=49.4$ efm$^{2}$ and 
$| G_{1}(\tfrac{1}{2}^{+},\tfrac{3}{2}^{+}) |=0.180$ $\mu_{N}$.}
\label{BEM7}
\vspace{10pt}
\begin{tabular}{crcrllll}
\hline
\noalign{\smallskip}
&& && \multicolumn{2}{c}{$B(E2)$} & \multicolumn{2}{c}{$B(M1)$} \\ 
$E_{\gamma}^{\exp}$ (keV) & $J'^{P'} \;$ & $\rightarrow$ & $J^P \;$ 
& Exp & Th & Exp & Th \\ 
\noalign{\smallskip}
\hline
\noalign{\smallskip}
2794 & $1/2^{+}$ & $\rightarrow$ & $3/2^{+}$ && 0 & 0.329(43)  & 0.032 \\ 
4684 & $3/2^{+}$ & $\rightarrow$ & $3/2^{+}$ && 0 & 0.0127(36) & 0.013$^{\ast}$ \\ 
4333 & $3/2^{+}$ & $\rightarrow$ & $5/2^{+}$ && 0 & 0.027(9)   & 0.019 \\
\noalign{\smallskip}
\hline
\end{tabular}
\end{table}

\subsection{$K^{P}=\tfrac{5}{2}^{+}(4526) \rightarrow K^{P}=\tfrac{5}{2}^{+}(4526)$}

$B(E2)$ and $B(M1)$ values for this band can be calculated using%
\begin{eqnarray}
B(E2;\tfrac{5}{2},J'^+ \rightarrow \tfrac{5}{2},J^+) &=& Q_{0}^{2} \frac{5}{16\pi} \left\langle J',\tfrac{5}{2},2,0|J,\tfrac{5}{2} \right\rangle^{2}
\nonumber\\
B(M1;\tfrac{5}{2},J'^+ \rightarrow \tfrac{5}{2},J^+) &=& 
\left\langle J',\tfrac{5}{2},1,0|J,\tfrac{5}{2} \right\rangle^2 
\left\vert G_{1}(\tfrac{5}{2}^{+}) \right\vert^2 ~.
\end{eqnarray}
Results are given in Table~\ref{BEM8}. No experimental data are available. 
$B(M1)$ values are calculated using an estimated value of 
$| G_{1}(\tfrac{5}{2}^{+}) |=1.45$ $\mu _{N}$.

\begin{table}
\centering
\caption[BEM8]{In-band $B(E2)$ values in e$^{2}$fm$^{4}$ and 
$B(M1)$ values in $\mu_{N}^{2}$ for the $K^{P}=\tfrac{5}{2}^{+}$(4525) band. 
The theoretical values were obtained with $Q_{0}=49.4$ efm$^{2}$ and 
$| G_{1}(5/2^{+}) |=1.45$ $\mu_N$.}
\label{BEM8}
\vspace{10pt}
\begin{tabular}{crcrllll}
\hline
\noalign{\smallskip}
&& && \multicolumn{2}{c}{$B(E2)$} & \multicolumn{2}{c}{$B(M1)$} \\ 
$E_{\gamma}^{\exp}$ (keV) & $J'^{P'} \;$ & $\rightarrow$ & $J^P \;$ 
& Exp & Th & Exp & Th \\ 
\noalign{\smallskip}
\hline
\noalign{\smallskip}
 905 & $ 7/2^+$ & $\rightarrow$ & $5/2^+$ && 86.7 && 0.45$^{\ast}$ \\ 
1123 & $ 9/2^+$ & $\rightarrow$ & $7/2^+$ && 73.6 && 0.65 \\ 
1686 & $11/2^+$ & $\rightarrow$ & $9/2^+$ && 56.6 && 0.76 \\ 
\nonumber\\
2028 & $ 9/2^+$ & $\rightarrow$ & $5/2^+$ && 24.3 &&  \\ 
2809 & $11/2^+$ & $\rightarrow$ & $7/2^+$ && 41.2 && \\
\noalign{\smallskip}
\hline
\end{tabular}
\end{table}

\subsection{$K^{P}=\tfrac{3}{2}^{+}(5549) \rightarrow K^{P}=\tfrac{3}{2}^{+}(5549)$ and 
$K^{P}=\tfrac{3}{2}^{+}(5826) \rightarrow K^{P}=\tfrac{3}{2}^{+}(5826)$}

$B(E2)$ and $B(M1)$ values can be calculated using Eqs.~(\ref{be33}) and 
(\ref{bm33}). The calculated values obtained with $Q_{0}=49.4$ efm$^{2}$ 
and $| G_{1}(\tfrac{3}{2}^{+}) |=0.85$ $\mu _{N}$, i.e. the same values as for the ground state band, are given in Table~\ref{BEM9}. No data are available for these bands.

\begin{table}
\centering
\caption[BEM9]{In-band $B(E2)$ values in e$^{2}$fm$^{4}$ and $B(M1)$ values in $\mu_{N}^{2}$ for the $K^{P}=\tfrac{3}{2}^{+}$(5549)/$K^{P}=\tfrac{3}{2}^{+}$(5826) bands. The theoretical values were obtained with $Q_{0}=49.4$ efm$^{2}$ and 
$| G_{1}(3/2^{+}) |=0.85$ $\mu_N$.}
\label{BEM9}
\vspace{10pt}
\begin{tabular}{crcrllll}
\hline
\noalign{\smallskip}
&& && \multicolumn{2}{c}{$B(E2)$} & \multicolumn{2}{c}{$B(M1)$} \\ 
$E_{\gamma}^{\exp}$ (keV) & $J'^{P'} \;$ & $\rightarrow$ & $J^P \;$ 
& Exp & Th & Exp & Th \\ 
\noalign{\smallskip}
\hline
\noalign{\smallskip}
224/ 348  & $5/2^{+}$ & $\rightarrow$ & $3/2^{+}$ && 83.2 && 0.19 \\ 
490/(238) & $7/2^{+}$ & $\rightarrow$ & $5/2^{+}$ && 52.0 && 0.26 \\ 
535/(630) & $9/2^{+}$ & $\rightarrow$ & $7/2^{+}$ && 34.0 && 0.29 \\
\noalign{\smallskip}
\hline
\end{tabular}
\end{table}

These bands are expected to decay to the ground state band in the same way
as the corresponding bands in $^{20}$Ne, $K^{P}=0^{+}(6.72)$ and 
$K^{P}=0^{+}(7.19)$, decay to the ground state band $K^{P}=0^{+}(0)$. The
transition moment for these decays in $^{20}$Ne is \cite{bijker3} 
$Q_{0t}=10.7$ efm$^{2}$. 
Out of band transitions $K^{P}=\tfrac{3}{2}^{+}(5549/5826) \rightarrow K^{P}=\tfrac{3}{2}^{+}(0)$ can be calculated using 
\begin{equation}
B(E2;\tfrac{3}{2},J'^+ \rightarrow \tfrac{3}{2},J^+) 
= \left( Q_{0t}\right)^{2} \frac{5}{16\pi} 
\left\langle J',\tfrac{3}{2},2,0|J,\tfrac{3}{2} \right\rangle^{2} ~,
\end{equation}
The value of the transition moment, $Q_{0t}$, is determined from the transition 
$B(E2;\tfrac{3}{2},\tfrac{5}{2}^+ \rightarrow \tfrac{3}{2},\tfrac{7}{2}^+)$ 
from the band $K^{P}=\tfrac{3}{2}^{+}(5826)$ to the ground state band to be $2.20$ efm$^{2}$. The calculated values for some decays of the two bands $K^{P}=\tfrac{3}{2}^{+}(5549)$ and $K^{P}=\tfrac{3}{2}^{+}(5826)$ to the ground
state band are shown in Table~\ref{BEM10}. The same transition moment $Q_{0t}$
appears to describe the decay of both bands.

\begin{table}
\centering
\caption[BEM10]{$K^{P}=\tfrac{3}{2}^{+}(5549)$/$K^{P}=\tfrac{3}{2}^{+}(5826)$ to $K^{P}=\tfrac{3}{2}^{+}(0)$ interband $B(E2)$ values in e$^{2}$fm$^{4}$. The theoretical values were obtained with $Q_{0t}=2.20$ efm$^{2}$.}
\label{BEM10}
\vspace{10pt}
\begin{tabular}{crcrlcc}
\hline
\noalign{\smallskip}
&&&&& $K^{P}=\tfrac{3}{2}^{+}$ & $K^{P}=\tfrac{3}{2}^{+}$ \\ 
$E_{\gamma}^{\exp}$ (keV) & $J'^{P'} \;$ & $\rightarrow$ & $J^P \;$ 
& Th & (5549) & (5826) \\ 
\noalign{\smallskip}
\hline
\noalign{\smallskip}
5549/5826 & $3/2^{+}$ & $\rightarrow$ & $3/2^{+}$ & 0.097 & 0.055(24) & \\ 
5198/5475 & $3/2^{+}$ & $\rightarrow$ & $5/2^{+}$ & 0.248 && \\ 
3804/4081 & $3/2^{+}$ & $\rightarrow$ & $7/2^{+}$ & 0.138 && \\ 
5773/6174 & $5/2^{+}$ & $\rightarrow$ & $3/2^{+}$ & 0.166 && \\ 
5422/5823 & $5/2^{+}$ & $\rightarrow$ & $5/2^{+}$ & 0.007 \\ 
4028/4429 & $5/2^{+}$ & $\rightarrow$ & $7/2^{+}$ & 0.138$^{\ast}$ && 0.138(172) \\ 
2906/3307 & $5/2^{+}$ & $\rightarrow$ & $9/2^{+}$ & 0.172 && \\
\noalign{\smallskip}
\hline
\end{tabular}
\end{table}

\section{Structure of $^{21}$Na}
\label{sec6}

We assume for $^{21}$Na a structure similar to $^{21}$Ne, with the odd
neutron replaced by an odd-proton, and perform an analysis of experimental
data \cite{firestone} similar to that of Sect.~\ref{sec3}.

\subsection{Assignments of states to bands}

\begin{figure}
\includegraphics[scale=0.8]{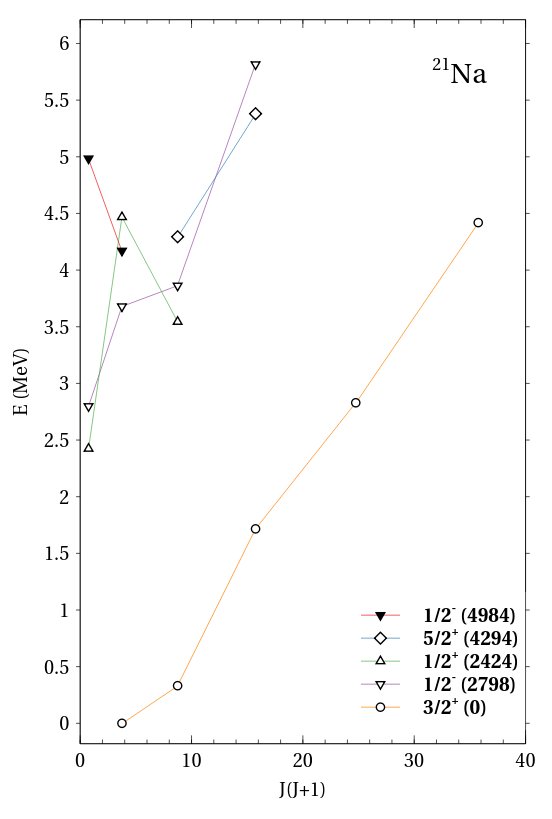}
\caption[Rotational bands in $^{21}$Na]
{Energies of assigned states in $^{21}$Na to rotational bands as a 
function of $J(J+1)$.}
\label{bandsNa}
\end{figure}

We are able to identify in this case four rotational bands for which the
values of $\varepsilon ,B$ and $a$ can be extracted, and fragments of two more bands as shown in Table~\ref{statesNa} and Fig.~\ref{bandsNa}.

\begin{table}
\centering
\caption{Rotational bands in $^{21}$Na}
\label{statesNa}
\vspace{10pt}
\begin{tabular}{ccrr}
\hline
\noalign{\smallskip}
$K^{P}(E_{\rm exc})$ & $J^{P}$ & $E_{\rm exp}$ & $E_{\rm th}$ \\ 
\noalign{\smallskip}
\hline
\noalign{\smallskip}
$3/2^{+}$(0) & $3/2^{+}$ & 0 & 0 \\ 
&   $5/2^{+}$  &   332  &  685 \\ 
&   $7/2^{+}$  &  1716  & 1644 \\ 
&   $9/2^{+}$  &  2829  & 2877 \\ 
&  $11/2^{+}$  & (4419) & 4380 \\ 
\noalign{\smallskip}
$1/2^{-}$(2798) & $1/2^{-}$ & 2798 & 2798 \\ 
&   $3/2^{-}$  &  3679 &  3679 \\ 
&   $5/2^{-}$  &  3862 &  3861 \\ 
&   $7/2^{-}$  &  5815 &  5916 \\ 
\noalign{\smallskip}
$1/2^{+}$(2424) & $1/2^{+}$ & 2424 & 2424 \\ 
& $3/2^{+}$ & 4468 & 4461 \\ 
& $5/2^{+}$ & 3544 & 3544 \\ 
\noalign{\smallskip}
$5/2^{+}$(4294) & $5/2^{+}$ & 4294 & 4294 \\ 
&   $7/2^{+}$ & (5380) & 5379 \\ 
\noalign{\smallskip}
$1/2^{-}$(4984) & $1/2^{-}$ & 4984 & \\ 
&   $3/2^{-}$  & 4170 & \\   
\noalign{\smallskip}
$1/2^{+}$(5457) & $1/2^{+}$ & 5457 & \\ 
\noalign{\smallskip}
\hline
\end{tabular}
\end{table}

\subsection{Cluster interpretation of rotational bands}

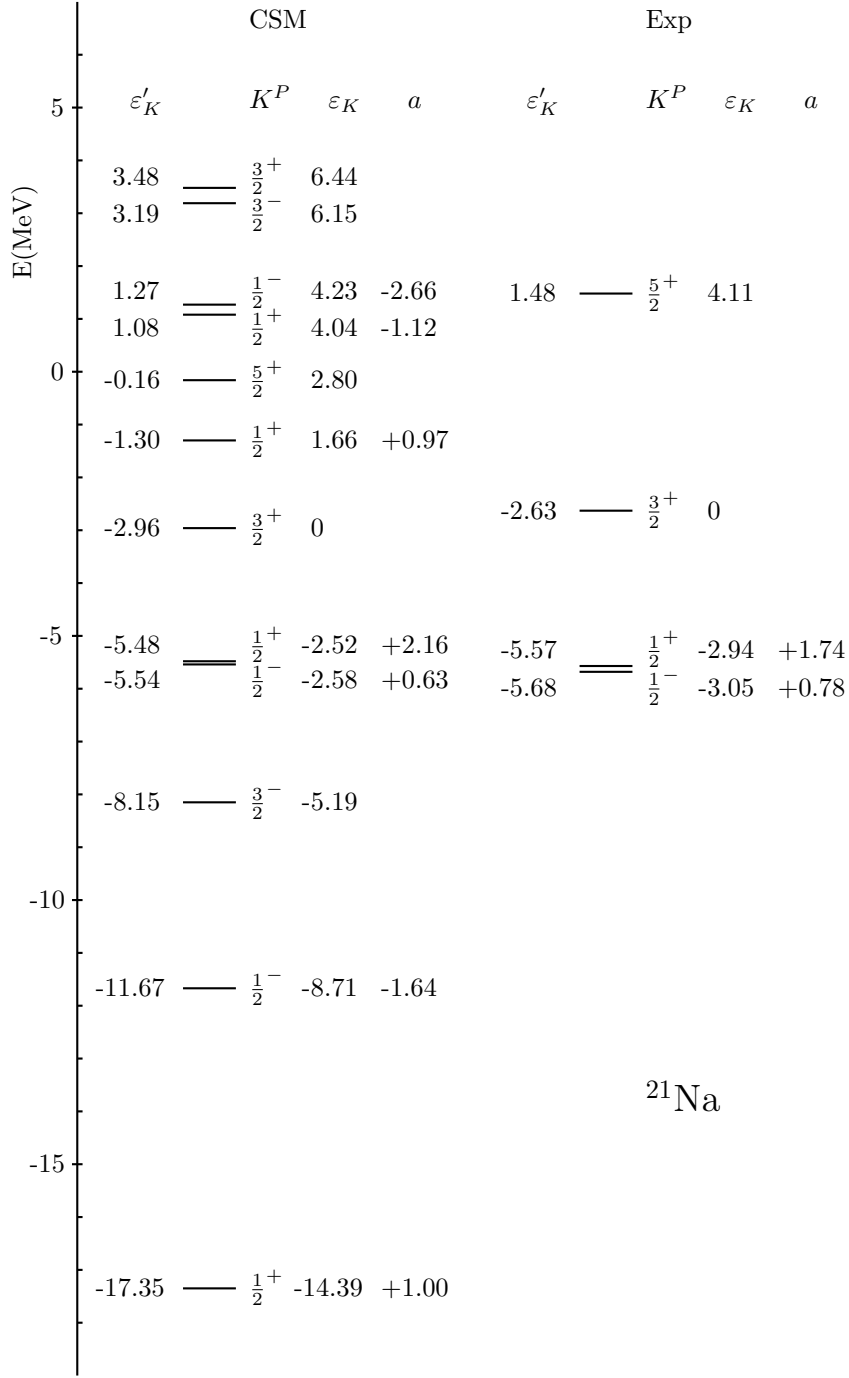
\begin{figure}
\centering
\setlength{\unitlength}{1pt}
\begin{picture}(350,520)(0,0)
\thicklines
\put ( 30,  0) {\line(0,1){520}}
\put ( 30, 20) {\line(1,0){2}}
\put ( 30, 40) {\line(1,0){2}}
\put ( 30, 60) {\line(1,0){2}}
\put ( 28, 80) {\line(1,0){4}}
\put ( 30,100) {\line(1,0){2}}
\put ( 30,120) {\line(1,0){2}}
\put ( 30,140) {\line(1,0){2}}
\put ( 30,160) {\line(1,0){2}}
\put ( 28,180) {\line(1,0){4}}
\put ( 30,200) {\line(1,0){2}}
\put ( 30,220) {\line(1,0){2}}
\put ( 30,240) {\line(1,0){2}}
\put ( 30,260) {\line(1,0){2}}
\put ( 28,280) {\line(1,0){4}}
\put ( 30,300) {\line(1,0){2}}
\put ( 30,320) {\line(1,0){2}}
\put ( 30,340) {\line(1,0){2}}
\put ( 30,360) {\line(1,0){2}}
\put ( 28,380) {\line(1,0){4}}
\put ( 30,400) {\line(1,0){2}}
\put ( 30,420) {\line(1,0){2}}
\put ( 30,440) {\line(1,0){2}}
\put ( 30,460) {\line(1,0){2}}
\put ( 28,480) {\line(1,0){4}}
\put ( 30,500) {\line(1,0){2}}
\put ( 12, 77) {-15}
\put ( 12,177) {-10}
\put ( 16,277) {-5}
\put ( 20,377) {0}
\put ( 20,477) {5}
\put (  5,415) {\rotatebox{90}{E(MeV)}}
\put( 70, 33.0) {\line(1,0){20}}
\put( 37, 30.0) {-17.35}
\put( 95, 30.0) {$\frac{1}{2}^+$}
\put(112, 30.0) {-14.39}
\put(145, 30.0) {+1.00}
\put( 70,146.6) {\line(1,0){20}}
\put( 37,143.6) {-11.67}
\put( 95,143.6) {$\frac{1}{2}^-$}
\put(115,143.6) {-8.71}
\put(145,143.6) {-1.64}
\put( 70,217.0) {\line(1,0){20}}
\put( 37,214.0) { -8.15}
\put( 95,214.0) {$\frac{3}{2}^-$}
\put(115,214.0) {-5.19}
\put( 70,269.2) {\line(1,0){20}}
\put( 37,260.2) { -5.54}
\put( 95,260.2) {$\frac{1}{2}^-$}
\put(115,260.2) {-2.58}
\put(145,260.2) {+0.63}
\put( 70,270.4) {\line(1,0){20}}
\put( 37,273.4) { -5.48}
\put( 95,273.4) {$\frac{1}{2}^+$}
\put(115,273.4) {-2.52}
\put(145,273.4) {+2.16}
\put( 70,320.8) {\line(1,0){20}}
\put( 37,317.8) { -2.96}
\put( 95,317.8) {$\frac{3}{2}^+$}
\put(115,317.8) {  0}
\put( 70,354.0) {\line(1,0){20}}
\put( 37,351.0) { -1.30}
\put( 95,351.0) {$\frac{1}{2}^+$}
\put(115,351.0) { 1.66}
\put(145,351.0) {+0.97}
\put( 70,376.8) {\line(1,0){20}}
\put( 37,373.8) { -0.16}
\put( 95,373.8) {$\frac{5}{2}^+$}
\put(115,373.8) { 2.80}
\put( 70,401.6) {\line(1,0){20}}
\put( 40,393.6) { 1.08}
\put( 95,393.6) {$\frac{1}{2}^+$}
\put(115,393.6) { 4.04}
\put(145,393.6) {-1.12}
\put( 70,405.4) {\line(1,0){20}}
\put( 40,407.4) { 1.27}
\put( 95,407.4) {$\frac{1}{2}^-$}
\put(115,407.4) { 4.23}
\put(145,407.4) {-2.66}
\put( 70,443.8) {\line(1,0){20}}
\put( 40,436.8) { 3.19}
\put( 95,436.8) {$\frac{3}{2}^-$}
\put(115,436.8) { 6.15}
\put( 70,449.6) {\line(1,0){20}}
\put( 40,450.6) { 3.48}
\put( 95,450.6) {$\frac{3}{2}^+$}
\put(115,450.6) { 6.44}
\put( 40,480.0) {$\quad \varepsilon'_K$}
\put( 95,480.0) {$K^P$}
\put(115,480.0) {$\quad \varepsilon_K$}
\put(145,480.0) {$\quad a$}
\put( 95,510.0) {CSM}
\put(220,266.4) {\line(1,0){20}}
\put(187,257.4) { -5.68}
\put(245,257.4) {$\frac{1}{2}^-$}
\put(265,257.4) {-3.05}
\put(295,257.4) {+0.78}
\put(220,268.6) {\line(1,0){20}}
\put(187,271.6) { -5.57}
\put(245,271.6) {$\frac{1}{2}^+$}
\put(265,271.6) {-2.94}
\put(295,271.6) {+1.74}
\put(220,327.4) {\line(1,0){20}}
\put(187,324.4) { -2.63}
\put(245,324.4) {$\frac{3}{2}^+$}
\put(265,324.4) {  0}
\put(220,409.6) {\line(1,0){20}}
\put(190,406.6) { 1.48}
\put(245,406.6) {$\frac{5}{2}^+$}
\put(265,406.6) { 4.11}
\put(190,480.0) {$\quad \varepsilon'_K$}
\put(245,480.0) {$K^P$}
\put(265,480.0) {$\quad \varepsilon_K$}
\put(295,480.0) {$\quad a$}
\put(245,510.0) {Exp}
\put(245,100.0) {\Large $^{21}$Na}
\end{picture}
\caption[Intrinsic energies in $^{21}$Na]{Energies of the intrinsic proton states in $^{21}$Na in the CSM (left) compared with the experimental intrinsic energies (right).}
\label{defspNa}
\end{figure}

The four observed rotational bands in $^{21}$Na are the mirror bands of the
four rotational bands $K^{P}=3/2^{+}(0)$, $1/2^{-}(2789)$, $1/2^{+}(2794)$ and $5/2^{+}(4526)$ in $^{21}$Ne. The fifth band 
is the mirror of the $K^{P}=1/2^{-}(5690)$ band in $^{21}$Ne. 
The sixth band does not have a direct counterpart in $^{21}$Ne. It is shown 
here because it may be evidence for the missing $K^{P}=1/2^{+}$ particle band in $^{21}$Ne with a tentative state at $5525$ keV. In order to analyze 
the band structure in $^{21}$Na we use the CSM Hamiltonian of Eq.~(\ref{hcsm}), 
to which the Coulomb interaction of the odd-proton with $^{20}$Ne is added 
\begin{equation}
H=\frac{\vec{p}^{2}}{2m}+V(\vec{r})+V_{so}(\vec{r})+V_{C}(\vec{r}).
\end{equation}%
The expression for the Coulomb interaction was given in Eq.~(32) of \cite{dellarocca1}. For simplicity in the numerical solution of the eigenvalue
problem, we approximate $V_{C}(\vec{r})$ with a form similar to $V(\vec{r})$ of 
Eq.~(\ref{vr}), but different values of $V_{1}$, $V_{2}$ and $\alpha_{1}$, 
$\alpha_{2}$, namely $V_{1C}=V_{2C}=1.71$ MeV and $\alpha_{1C}=\alpha_{2C}=0.034$ 
fm$^{-2}$. The single-particle intrinsic levels with the Coulomb interaction
included are given in the left side of Fig.~\ref{defspNa}. 

\begin{figure}
\centering
\setlength{\unitlength}{1pt}
\begin{picture}(230,285)(-10,-30)
\thicklines
\put ( 30,-30) {\line(0,1){285}}
\put ( 28, 30) {\line(1,0){4}}
\put ( 30, 60) {\line(1,0){2}}
\put ( 30, 90) {\line(1,0){2}}
\put ( 30,120) {\line(1,0){2}}
\put ( 30,150) {\line(1,0){2}}
\put ( 28,180) {\line(1,0){4}}
\put ( 30,210) {\line(1,0){2}}
\put ( 30,240) {\line(1,0){2}}
\put ( 15, 27) {0}
\put ( 15,177) {5}
\put (-10,210) {E(MeV)}
\put( 40, 30.0) {\line(1,0){15}}
\put( 40, 39.9) {\line(1,0){15}}
\put( 40, 81.5) {\line(1,0){15}}
\put( 40,114.9) {\line(1,0){15}}
\put( 40,162.6) {\line(1,0){15}}
\put( 35,-10.0) {$\begin{array}{c} \frac{3}{2}^+ \\ \\ 137 \\ \mbox{} \end{array}$}
\put( 57, 25.0) {$\frac{3}{2}^+$}
\put( 57, 38.9) {$\frac{5}{2}^+$}
\put( 57, 78.5) {$\frac{7}{2}^+$}
\put( 57,111.9) {$\frac{9}{2}^+$}
\put( 57,159.6) {$\frac{11}{2}^+$}
\put( 80,114.0) {\line(1,0){15}}
\put( 80,140.4) {\line(1,0){15}}
\put( 80,145.8) {\line(1,0){15}}
\put( 80,204.3) {\line(1,0){15}}
\put( 75,-10.0) {$\begin{array}{c} \frac{1}{2}^- \\ \\ 165 \\ 0.78 \end{array}$}
\put( 97,111.0) {$\frac{1}{2}^-$}
\put( 97,133.4) {$\frac{3}{2}^-$}
\put( 97,146.8) {$\frac{5}{2}^-$}
\put( 97,201.3) {$\frac{7}{2}^-$}
\put(120,102.6) {\line(1,0){15}}
\put(120,136.2) {\line(1,0){15}}
\put(120,164.1) {\line(1,0){15}}
\put(115,-10.0) {$\begin{array}{c} \frac{1}{2}^+ \\ \\ 248 \\ 1.74 \end{array}$}
\put(137, 99.6) {$\frac{1}{2}^+$}
\put(137,133.2) {$\frac{5}{2}^+$}
\put(137,161.1) {$\frac{3}{2}^+$}
\put(160,158.7) {\line(1,0){15}}
\put(160,191.4) {\line(1,0){15}}
\put(155,-10.0) {$\begin{array}{c} \frac{5}{2}^+ \\ \\ 155 \\ \mbox{} \end{array}$}
\put(177,155.7) {$\frac{5}{2}^+$}
\put(177,188.4) {$\frac{7}{2}^+$}
\end{picture}
\caption{Cluster interpretation of the rotational bands in $^{21}$Na. The
bands are labeled by $K^{P}$ and the values of the rotational
parameter, $B$, and decoupling parameter, $a$.}
\label{Na21}
\end{figure}
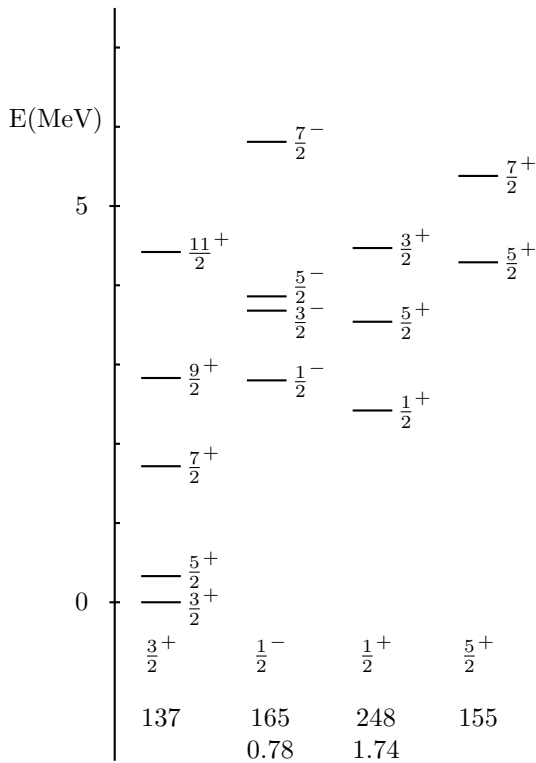

We calculate the proton
separation energy $S_{p}($calc$)=2.75$ MeV to be compared with the
experimental value $S_{p}(\exp )=2.42168(28)$ MeV. This is a remarkable
result since there are no parameters in the calculation, the strength of the
Coulomb interaction being fixed by the charge of $^{20}$Ne, $Z=10$. Using
Eq.~(\ref{erot}) we can extract the values of $\varepsilon$, $B$ and $a$ for each of the four observed bands, a summary of which is given in Table~\ref{assignmentsNa}. In Fig.~\ref{Na21} we show the corresponding level scheme.

\begin{table}
\centering
\caption[Assignments]{Summary of assignments in $^{21}$Na}
\label{assignmentsNa}
\vspace{10pt}
\begin{tabular}{lccccccc}
\hline
\noalign{\smallskip}
$\#$ & $K^P$ & $E$(keV) & $B$(keV) & $a$ & $\varepsilon$(keV) 
& $\varepsilon_K$(MeV) & $\varepsilon'_K$(MeV) \\
\noalign{\smallskip}
\hline
\noalign{\smallskip}
1 (p) & $3/2^{+}$ &    0 & 137(4)  &         & --206(21) &   0    & --2.63 \\
2 (h) & $1/2^{-}$ & 2798 & 165(12) & 0.78(6) &  2845(5)  & --3.05 & --5.68 \\
3 (h) & $1/2^{+}$ & 2424 & 248(16) & 1.74(8) &  2731(30) & --2.94 & --5.57 \\
4 (p) & $5/2^{+}$ & 4294 & 155(12) &         &  3907(39) &   4.11 &   1.48 \\
\noalign{\smallskip}
\hline
\end{tabular}
\end{table}

The cluster interpretation of the four observed bands in $^{21}$Na is
similar to that of their mirror bands in $^{21}$Ne and we do not repeat it
here. We consider instead the Coulomb displacement energies $\Delta E_{K}$
of the four observed bands in $^{21}$Na. In Table~\ref{cde} we compare the
calculated values with the experimental values.

\begin{table}
\centering
\caption[Coulomb displacement energies]
{Coulomb displacement energies for $^{21}$Ne$ - ^{21}$Na. All values in MeV.}
\label{cde}
\vspace{10pt}
\begin{tabular}{ccrrrrrr}
\hline
\noalign{\smallskip}
$\#$ & $K^P$ & \multicolumn{2}{c}{$\varepsilon'_K(\rm th)$} 
& $\Delta \varepsilon'_K(\rm th)$ & \multicolumn{2}{c}{$\varepsilon'_K(\rm exp)$} 
& $\Delta \varepsilon'_K(\rm exp)$ \\
&& $^{21}$Ne & $^{21}$Na && $^{21}$Ne & $^{21}$Na & \\ 
\noalign{\smallskip}
\hline
\noalign{\smallskip}
1 (p) & $3/2^{+}$ &  --7.22 & --2.96 & 4.26 &  --6.97 & --2.63 & 4.34(2) \\
2 (h) & $1/2^{-}$ & --10.45 & --5.54 & 4.91 &  --9.96 & --5.68 & 4.28(4) \\
3 (h) & $1/2^{+}$ &  --9.84 & --5.48 & 4.36 & --10.28 & --5.57 & 4.71(4) \\
4 (p) & $5/2^{+}$ &  --4.37 & --0.16 & 4.21 &  --2.56 &   1.48 & 4.04(2) \\
\noalign{\smallskip}
\hline
\end{tabular}
\end{table}

While the CSM correctly describes the Coulomb displacement energies of the
particle states $K^{P}=3/2^{+}$ and $5/2^{+}$, it fails 
in describing the displacement energies of the hole states 
$K^{P}=1/2^{-}$ and $1/2^{+}$. This failure may be due to the neglect of residual interactions in the single particle approach of this paper. While the particle states in $^{21}$Na can be associated with the cluster 
structure $^{20}$Ne+p, the hole states are associated with the $^{19}$F+2p cluster
structure. The latter structure has an additional proton-proton interaction
not included in the CSM.

\subsection{Electromagnetic transition rates}

The available experimental information in $^{21}$Na is not as extensive as
in $^{21}$Ne. $B(\lambda )$ values and electric and magnetic moments can be
calculated still as in Eqs.~(\ref{blam}-\ref{qmu}). However, in $^{21}$Na the 
single particle is a proton and thus there is a contribution to electric
transitions.

\subsubsection{$K^{P}=\tfrac{3}{2}^{+}(0) \rightarrow K^{P}=\tfrac{3}{2}^{+}(0)$}

Some transition rates have been measured for this band. The $B(E2)$ values
and quadrupole moments $Q^{(2)}$ can be calculated as in Eq.~(\ref{be33}) but
with $Q_{0}=Q_{0c}+Q_{0p}$. From the $B(E2;\tfrac{3}{2},\tfrac{5}{2}^+ \rightarrow \tfrac{3}{2},\tfrac{3}{2}^+)$ we extract the value of $Q_{0}$ in $^{21}$Na to be $Q_{0}(^{21}$Na$)=62.7(20)$ efm$^{2}$. Using the value of $Q_{0c}(^{21}$Ne$)=49.4(20)$ efm$^{2}$, we obtain $Q_{0p}=13.3(20)$ efm$^{2}$. 
With the value of $Q_{0}=62.7$ efm$^{2}$ we calculate all $B(E2)$ values and quadrupole moments as given in Tables~\ref{BEM11} and \ref{mom2}.

\begin{table}
\centering
\caption[BEM11]{In-band $B(E2)$ values in e$^{2}$fm$^{4}$ and 
$B(M1)$ values in $\mu_{N}^{2}$ for the $K^{P}=\tfrac{3}{2}^{+}(0)$ band. 
The theoretical values were obtained with $Q_{0}=62.7$ efm$^{2}$ and $\left| G_{1}(\tfrac{3}{2}^{+}) \right|=1.0$ $\mu_N$.}
\label{BEM11}
\vspace{10pt}
\begin{tabular}{crcrllll}
\hline
\noalign{\smallskip}
&& && \multicolumn{2}{c}{$B(E2)$} & \multicolumn{2}{c}{$B(M1)$} \\ 
$E_{\gamma}^{\exp}$ (keV) & $J'^{P'} \;$ & $\rightarrow$ & $J^P \;$ 
& Exp & Th & Exp & Th \\ 
\noalign{\smallskip}
\hline
\noalign{\smallskip}
 332 & $ 5/2^{+}$ & $\rightarrow$ & $ 3/2^{+}$ 
& 134.2(103) & 134.0$^{\ast}$ & 0.151(2) & 0.27 \\ 
1384 & $ 7/2^{+}$ & $\rightarrow$ & $ 5/2^{+}$ 
& 55.0(270) & 83.7 & 0.358(89) & 0.36$^{\ast}$ \\ 
\noalign{\smallskip}
1716 & $ 7/2^{+}$ & $\rightarrow$ & $ 3/2^{+}$ & 72.3(275) & 55.9 & & \\ 
\noalign{\smallskip}
\hline
\end{tabular}
\end{table}

\begin{table}
\centering
\caption[Moments]{Spectroscopic quadrupole moment in efm$^2$ and magnetic moment 
in $\mu_N$ of the $J^P=\tfrac{3}{2}^{+}$ ground state.}
\label{mom2}
\vspace{10pt}
\begin{tabular}{cccc}
\hline
\noalign{\smallskip}
& Exp & Th & \\ 
\noalign{\smallskip}
\hline
\noalign{\smallskip}
  $Q^{(2)}(\tfrac{3}{2},\tfrac{3}{2}^+)$ & $+12.4(14)$ & $+12.6$ & efm$^{2}$ \\
\noalign{\smallskip}
$\mu^{(1)}(\tfrac{3}{2},\tfrac{3}{2}^+)$ & $+2.38630(10)$ & $+1.84$ & $\mu_N$ \\
\noalign{\smallskip}
\hline
\end{tabular}
\end{table}

Similarly, the $B(M1)$ values and magnetic moments are given by Eq.~(\ref{bm33}. From the $B(M1;\tfrac{3}{2},\tfrac{7}{2}^+ \rightarrow \tfrac{3}{2},\tfrac{5}{2}^+)$ we extract the value $\left\vert G_{1}(3/2^{+})\right\vert =1.0(3)$ $\mu_{N}$. With this value we calculate the $B(M1)$ values given in Table~\ref{BEM11}. Using $G_{1}(3/2^{+})=1.0$ $\mu _{N}$ and adding $G_{1R}=0.50$ $\mu_{N}$ we obtain
the magnetic moment given in Table~\ref{mom2}. While the electric transitions and
quadruple moment appear to be well described by the cluster model, the
magnetic moment is underestimated by 20 \%.

\section{Summary and conclusions}
\label{sec7}

In this article, we have investigated the structure of $^{21}$Ne and its
mirror nucleus $^{21}$Na, and analyzed it in terms of the cluster shell
model (CSM) \cite{dellarocca1}. The structure of these nuclei appears to be
rather complex with three types of rotational bands, particle bands, hole
bands and vibrational bands. In $^{21}$Ne, three particle bands, 
$K^{P}=3/2^{+}(0)$, $5/2^{+}(4526)$ and  
$1/2^{-}(5690)$, two hole bands, $K^{P}=1/2^{-}(2789)$ and $1/2^{+}(2794)$, and two vibrational bands, $K^{P}=3/2^{+}(5549)$ and $3/2^{+}(5826)$ have been identified, while in $^{21}$Na only two particle bands, $K^{P}=3/2^{+}(0.0)$ and $5/2^{+}(4294)$, 
and two hole bands, $K^{P}=1/2^{-}(2798)$ and $1/2^{+}(2424)$ have been clearly identified, together with some fragments of two bands with $K^{P}=1/2^{-}(4984)$ and $1/2^{+}(5457)$.
The CSM appears to describe most of the observed properties of these bands
well, with the exception of the non-occurrence of a low-lying $K^P=1/2^{+}$ particle band in addition to the observed $K^{P}=1/2^{+}(2794)$ hole band 
and of the Coulomb displacement energy of the hole bands. The observed
properties of the rotational bands in $^{21}$Ne and $^{21}$Na support the
cluster interpretation of the particle bands as $^{20}$Ne+n and $^{20}$Ne+p
and of the hole bands as $^{19}$Ne+2n and $^{19}$F+2p.

States in a bi-pyramidal potential have features similar to those of states
in an ellipsoidal potential (Nilsson model). This is due to the fact that
the bi-pyramid can be inscribed into an ellipsoid. The Nilsson model gives
therefore a description of the observed single particle states in $^{21}$Ne
as good as the cluster model. As discussed in \cite{bijker5} the main
difference between the cluster \cite{brink} and the quadrupole collective 
\cite{bohr} description of $^{20}$Ne is in the corresponding vibrational
spectra. For a bi-pyramid one has 9 vibrations, 3 singly degenerate and 3
doubly degenerate, while for the ellipsoid one has three vibrations, one
singly degenerate ($\beta $-vibration) and one doubly degenerate ($\gamma $%
-vibration). In $^{21}$Ne one therefore expects a difference between
vibrational states for the cluster and the Nilsson model. Two of these
vibrational states have been clearly identified, $K^{P}=3/2^{+}(5549)$ 
and $3/2^{+}(5826)$, pointing out to the cluster interpretation, since in
the Nilsson interpretation one expects only one band with $K^{P}=3/2^{+}$ ($\beta $-vibration). Additional vibrational bands appear to be present at
higher excitation energy $\geq 6$ MeV with $K^{P}=1/2^{-}$ and $3/2^{-}$, but their identification is very difficult due to the high density of states at this excitation energy.

Our assignment of states into bands has relied mainly on energies and
electromagnetic transition rates. Our identification as particle or hole
states has relied on intensities of $^{20}$Ne$(d,p) ^{21}$Ne (for particle
states) and $^{22}$Ne$(p,d) ^{21}$Ne reactions (for hole states) \cite{howard1}. 
Our assignments agree only in part with those of \cite{wheldon}. These
authors use the incomplete fusion reaction $^{16}$O($^{7}$Li,$pn) ^{21}$Ne to
study states in $^{21}$Ne. It appears that this reaction populate strongly
the ground state band $K^{P}=3/2^{+}$ and the two hole bands,  
$K^{P}=1/2^{-}$ and $1/2^{+}$, and weakly the states $J^{P}=3/2^{-}$ at $4723$ keV, $9/2^{(-)}$ at $6642$ keV and $(7/2^{-})$ at $7370$ keV. In the incomplete fusion reaction, 2 protons and 3 neutrons are transfered. If we decompose this transfer into $^{4}$He+n and $^{3}$He+2n we can reconcile the fusion reaction results with our and Nilsson assignments into particle states and hole states.

Finally, the analysis of the rotational bands of $^{21}$Ne, $^{21}$Na
presented here can also provide the ground for testing microscopic theories
of light nuclei, such as the SDPF interaction model of \cite{caurier}, in
particular the extent to which cluster features can be obtained from large
scale shell model calculations. In this context, also of importance it would
be a large scale shell model calculation of $^{21}$Na to see the extent to
which the microscopic shell model can describe Coulomb dispacement energies.

\section{Acknowledgements}

This work was supported in part under research grant IN101320 from
PAPIIT-DGAPA. We wish to thank D. Mengoni for stimulating the study of $^{21}$Ne following an experiment performed at LNL in Legnaro, Italy, and L.
Fortunato and E. Buonocore for correspondence in the early stages of this
investigation.

\end{document}